\documentclass[12pt]{article}
\pdfoutput=1
\pdfpageattr{/Group << /S /Transparency /I true /CS /DeviceRGB>>} 
\usepackage[nottoc]{tocbibind}
\usepackage{graphicx}
\usepackage{amsmath}
\usepackage{amsfonts}
\usepackage{amssymb}
\usepackage{setspace}
\usepackage{subcaption}
\usepackage{caption}
\usepackage[hidelinks]{hyperref}
\usepackage{mathrsfs}
\usepackage{enumerate}
\usepackage[utf8]{inputenc}
\numberwithin{equation}{section}
\usepackage[normalem]{ulem}
\usepackage{color}

\def\e{{\rm e}}

\def\eps{\epsilon}
\def\d{\partial}
\def\l{\left(}
\def\r{\right)}

\newcommand{\be}{\begin{equation}}
\newcommand{\ee}{\end{equation}}
\newcommand{\bea}{\begin{eqnarray}}
\newcommand{\eea}{\end{eqnarray}}
\newcommand{\bg}{\begin{gather}}
\newcommand{\eg}{\end{gather}}
\newcommand{\bseq}{\begin{subequations}}
\newcommand{\eseq}{\end{subequations}}

\newcommand{\comment}[1]{}
\newcommand{\expect}[1]{\left\langle #1 \right\rangle}

\def\ep{\epsilon}
\def\vep{\varepsilon}

\def\d{\partial}

\begin{document}

\vspace*{-1. cm}
\begin{center}

{\bf \Large Asymptotic Fragility, Near $AdS_2$ Holography and  $T\bar{T}$} \\

\vskip 1cm
{\normalsize  {Sergei Dubovsky$^{a,b}$, Victor Gorbenko$^c$, and Mehrdad Mirbabayi$^{c,d}$}}

\vskip 0.5cm

{\normalsize $^{a}$ {\em Center for Cosmology and Particle Physics, New York University, New York, NY, USA}}

{\normalsize $^{b}$ {\em Perimeter Institute for Theoretical Physics, Waterloo, ON, Canada}}

{\normalsize $^c$ {\em Stanford Institute for Theoretical Physics, Stanford University, Stanford, CA, USA}}

{\normalsize $^d$ {\em Abdus Salam International Centre for Theoretical Physics, Trieste, Italy}}
\end{center}

\vspace{.8cm}

\hrule \vspace{0.3cm}
\begin{center}
{\small  \noindent \textbf{Abstract} }
\\[0.3cm]
\end{center}
\noindent We present the exact solution for the scattering problem in the flat space Jackiw--Teitelboim (JT) gravity
coupled to an arbitrary quantum field theory. JT gravity results in a gravitational dressing of field theoretical scattering amplitudes. The exact expression for the dressed $S$-matrix was previously known as a solvable example of a novel UV asymptotic behavior, dubbed asymptotic fragility. This dressing is equivalent to the $T\bar{T}$ deformation
of the initial quantum field theory. JT gravity coupled to a single massless boson  provides a promising 
action formulation for an integrable approximation to  the worldsheet theory of confining strings in 3D gluodynamics.
We also derive the dressed $S$-matrix as a flat space limit of the near $AdS_2$ holography.
We show that in order to preserve the flat space unitarity the conventional Schwarzian dressing of boundary correlators needs to be slightly extended. Finally, we propose a new simple expression for flat space amplitudes of massive particles in terms of correlators of holographic CFT's.

\vspace{0.3cm}
\hrule

\begin{flushleft}
\end{flushleft}
\newpage
\vspace{-1cm}
\tableofcontents
\section{Introduction and Summary}
Physics in two space-time dimensions provides a useful playground to test ideas about strongly coupled quantum field theory
and gravity in a safe child proof environment. In spite of a long history, studies of two-dimensional systems continue to provide
new insights. The current study is triggered by the following three recent developments in the two-dimensional world.
The earliest one is the gravitational dressing introduced in  \cite{Dubovsky:2012wk,Dubovsky:2013ira}, which is an explicit $S$-matrix construction of UV complete quantum theories exhibiting a novel type of high energy behavior, dubbed asymptotic fragility.
The second is the recent progress in resolving the puzzles of $AdS_2/CFT_1$ correspondence, resulting in the Schwarzian descritption of the $NAdS_2$ holography (where $N$ stands for {\it Near}) \cite{Almheiri:2014cka,Kitaev,Sachdev:1992fk,Maldacena:2016hyu,Jensen:2016pah,Maldacena:2016upp,Engelsoy:2016xyb,Cvetic:2016eiv}. Last but not least, a solvable {\it irrelevant} deformation by the $T\bar{T}$ operator, present in any 
local two-dimensional quantum field theory, has been identified recently in \cite{Smirnov:2016lqw,Cavaglia:2016oda}.

The main goal of the present paper is to argue that these three developments are very closely related. Namely, the gravitational dressing procedure  is the $S$-matrix definition of the $T\bar{T}$ deformation. The resulting
theory can be obtained as a flat space limit of $NAdS_2$ holography and provides a solution to the flat space Jackiw--Teitelboim (JT) gravity \cite{Jackiw:1984je,Teitelboim:1983ux}.

In the reminder of this introductory section we present a lightning review of gravitational dressing, $NAdS_2$ holography and $T\bar{T}$ deformation and present heuristic arguments indicating that the three are closely related. In the rest of the paper we present more detailed and solid (to the extent we succeded) arguments for the above equivalence.

\begin{figure}[t!]
  \begin{center}
        \includegraphics[height=8cm]{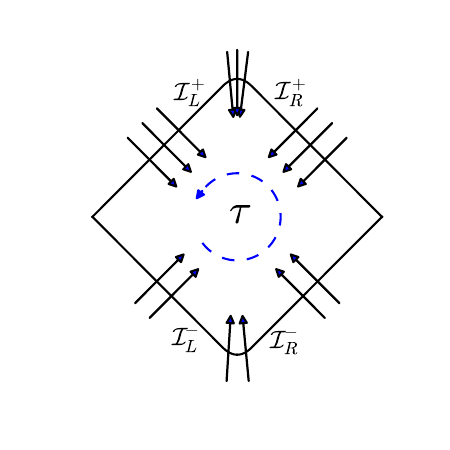} 
           \caption{ Particles at the asymptotic boundary of Minkowski space. Arrows correspond to on-shell two-momenta. $\tau$ is the coordinate parametrizing the boundary and ordering is given by the rapidities as described in the text. }
        \label{fig_dressing}
    \end{center}
\end{figure}

The gravitational dressing operates as follows. Let us start with an arbitrary quantum field theory in two dimensions, which can be defined by its $S$-matrix elements $S(\{p_i\})$, where all momenta are taken to be incoming.
Then the gravitationally dressed ${S}$-matrix is defined as
\be
\label{dressing}
\hat{S}(\{p_i\})=S(\{p_i\})e^{{i\ell^2\over 4}\sum_{i<j}p_i*p_j}\;,
\ee
where $\ell^2$ is a parameter characterizing the dressing,
\[
p_i*p_j=\epsilon_{\alpha\beta}p_i^{\alpha}p_j^{\beta}\;,\qquad  \ep^{01} =-\ep^{10}= 1,
\]
and the momenta are ordered according to the corresponding rapidities $\beta_i$: $i<j$ if $i$ and $j$ are both incoming particles and $\beta_i> \beta_j$, or if they are both outgoing and $\beta_i < \beta_j$, or else if $i$ is incoming and $j$ outgoing, see Fig.~\ref{fig_dressing}. The claim is that dressed amplitudes satisfy all the requirements expected for a healthy $S$-matrix. Their high energy behavior, however, is incompatible with the existence of a UV fixed point. Instead, a dressed theory exhibits many features expected from a gravitational theory, rather than from a conventional quantum field theory. In particular, as discussed in detail in  \cite{Dubovsky:2012wk}, one does not expect to find sharply defined local observables in a dressed theory.
The first and, perhaps, the most important example of dressing is provided by the worldsheet theory of an infinitely long critical (super)string, which in the Polyakov formalism indeed takes the form of a two-dimensional gravity.

The dressing formula (\ref{dressing}) was obtained as a result of a guesswork. One of our results here is to provide a systematic derivation for the dressing.
It is useful to have in mind the following path integral presentation for the dressed $S$-matrix,
\be
\hat{S}(\{p_i\})=\int {\cal D}X^\alpha S(\{p_i\})e^{iS_{CS}[X^\alpha]+{ i\sum_ip_{i\alpha}X^\alpha(\tau_i)}}\;,
\label{dressFI}
\ee
where
\be
S_{CS}= \ell^{-2}\oint d\tau\eps_{\alpha\beta}X^\alpha\d_\tau X^\beta
\label{CS}    
\ee
is the action of a Chern--Simons quantum mechanics, which can be thought of as living at the boundary of the Minkowski space-time. This presentation is inspired by the derivation of the eikonal scattering in higher-dimensional gravity presented in \cite{Verlinde:1991iu} and by the similarity between the dressing phases in (\ref{dressing}) and the gravitational eikonal phase (for
massless scattering they are exactly the same).

Let us turn now to the $NAdS_2$ holography.
The  results of \cite{Jensen:2016pah,Maldacena:2016upp,Engelsoy:2016xyb}, relevant for the present discussion, can be summarized as follows. Let us start with a quantum field theory in the rigid $AdS_2$ space. Then one may define a set of boundary correlators. 
The effect of introducing dynamical  gravity in the bulk with the action of the JT form,
\be
\label{JT}
S_{JT}=\int d^2\sigma\sqrt{-g}
 \l\phi\l R+{2\over L^2}\r-\Lambda\r  + 2\phi_b\oint_{\cal C} du\sqrt{g_{uu}} K
\ee
can be described by the following dressing formula
\be
\label{AdSdressing}
\langle  {\cal V}_1\l u_1 \r\ldots {\cal V}_n\l u_n \r\rangle_{JT}=\int{\cal D}t e^{{ (\Lambda L^2-2\phi_b)
\oint_{\cal C} d u Sch(t(u))}}\prod_i\l t'(u_i)^{\Delta_i}\r\langle  {\cal V}_1\l t(u_1)\r\ldots {\cal V}_n\l t(u_n)\r\rangle\;, 
\ee
where the Schwarzian $Sch(t(u))$ is defined as
\be\label{Sch}
Sch(t(u)) = \l\frac{t''}{t'}\r' -\frac{1}{2} \l\frac{t''}{t'}\r^2\;.
\ee
Note, that our notations here are different from the rest of $NAdS_2$ literature.
Conventionally, the vacuum energy $\Lambda$ is set to zero, and the dressing is parametrized by the boundary value of the dilaton $\phi_b$. As explained in more detail in section~\ref{sec:schwarz}, the two descriptions are equivalent, because at finite $AdS_2$ length $L$ the vacuum energy can be traded for $\phi_b$
 by performing a constant field shift $\phi\to\phi+const$.
We keep $\Lambda$ in (\ref{AdSdressing}) and will set 
\be
\phi_b =0\;,
\ee
because eventually we are interested in the flat limit $L\to\infty$, where such a redefinition is impossible. An interesting aspect of the expression (\ref{AdSdressing}) is that the dressed correlators 
are not conformally invariant. This is the reason one refers to this setup as $NAdS_2$ holography, rather than $AdS_2/CFT_1$ correspondence.

Finally, the $T\bar{T}$ deformation operates as follows \cite{Smirnov:2016lqw,Cavaglia:2016oda} (a 3D holographic picture of this deformation was proposed in \cite{McGough:2016lol}). Based on the special
remarkable properties \cite{Zamolodchikov:2004ce} of the operator\footnote{For a conformal theory this operator reduces to the product of holomorphic and antiholomorphic components of the energy-momentum tensor $T\bar{T}$, which explains its name. Away from a fixed point the operator is equal to $T\bar{T}-{1\over 4}\Theta^2$, where $\Theta$ is the trace of the energy-momentum tensor.}
\[
T\bar{T}\equiv {1\over 2}\l T_{\alpha\beta}T^{\alpha\beta}-T_\alpha^{\alpha2}\r
\]     
it was observed that an {\it arbitrary} quantum field theory deformed by $T\bar{T}$ gives rise to an RG trajectory, which is exactly solvable in the following sense. If $E_0(n,P,R)$ is the energy spectrum of the undeformed theory put on a circle of size $R$, with $P$ being a spatial momentum, then the finite volume spectrum of the deformed theory is given by the solution of the following partial differential equation
\be
\label{Burgers}
\d_\lambda E_\lambda (n,P,R)=E_\lambda (n,P,R)\d_RE_\lambda(n,P,R)+{P(R)^2\over R}\;.
\ee
Here $\lambda$ is the parameter of the deformation, so that the undeformed spectrum serves as the initial condition at $\lambda=0$.
Note that the unusual property of this deformation is that $T\bar{T}$ is an irrelevant  operator, so the RG trajectory shoots from IR into UV.

As already said, the principal claim of the current paper is that the gravitational dressing and the $T\bar{T}$ deformation are two different descriptions of the same RG flow, which can be described also by coupling the undeformed theory to the flat space 
({\it i.e.}, $L\to\infty$) JT gravity. The parameters of the three constructions are identified as\footnote{The relation of $\lambda$ to the deformation parameters $\alpha$ and $t$ of \cite{Smirnov:2016lqw} and \cite{Cavaglia:2016oda} is $\lambda=-\alpha=t$}
\[
 \ell^{-2}=-{\Lambda\over 2}={1\over 2}\lambda^{-1}\;.
\] 

Let us explain why this result is natural to expect. The similarity between the flat space gravitational dressing and the Schwarzian dressing in $NAdS_2$ holography is manifest. In both cases all the effects of gravity can be described by introducing a coupling between the non-gravitational asymptotic observables---be it $S$-matrix elements or boundary correlators--- and a boundary quantum mechanics.
Furthermore, it is easy to see that the boundary Chern--Simons theory { (\ref{CS})} arises naturally in the JT gravity. Indeed, in the bulk the JT dilaton $\phi$ plays the role of a Lagrange multiplier, ensuring that the metric is flat (at $L=\infty$). Hence, at $L=\infty$  
the path integral describing a matter system coupled to the JT gravity can be written schematically as 
\be
Z=\int{\cal D}X^a{\cal D}\psi e^{i\int d^2\sigma\sqrt{-g_f}\l-\Lambda+{\cal L}_m(\psi,g_f)\r}\;,
\ee
where ${\cal L}_m$ is the matter Lagrangian and $g_f$ is a general  flat  metric, which can be presented as
\[
 g_{f\alpha\beta}=\d_\alpha X^a\d_\beta X^b\eta_{ab}\;.
\]
Hence, the vacuum energy term turns into the action of a topological theory; when integrated by parts the latter is exactly the boundary Chern--Simons quantum mechanics (\ref{CS}) with 
\be
\label{ellL}
\ell^{-2}=-{\Lambda\over 2}\;.
\ee
Of course, this heuristic argument falls short of the derivation that the flat space JT gravity results in the gravitationally dressed amplitudes.
In section~\ref{sec:solution} we will present an actual proof that this is the case. We will show how the dressing phase shift arises
from the JT action  perturbatively at the leading order in $\ell^2$ expansion, as well as present a non-perturbative derivation of the full dressing phase.

Let us turn now to the $T\bar{T}$ deformation. There are several pieces of evidence strongly indicating that it is equivalent to the gravitational dressing. In the very first example of dressing---the critical string---the finite volume spectrum is known exactly and coincides with the one given by the $T\bar{T}$ deformation, as first pointed out in \cite{Caselle:2013dra}. This agreement extends to deformations of conformal field theories (CFT) and of integrable models, where the effect of dressing on the finite volume spectrum can be described  through the Thermodynamic Bethe Ansatz (TBA), see \cite{Smirnov:2016lqw,Cavaglia:2016oda} for the explicit discussion in the sine-Gordon case. In addition, as we will see in section~\ref{sec:solution}, the $T{\bar T}$ operator arises at the leading order in perturbation theory based on the JT action for an arbitrary matter sector.

It appears that the JT gravity provides a tractable path integral formulation for the deformation, which should allow to directly calculate the deformed finite volume spectrum  beyond the perturbation theory. Unfortunately, due to several (hopefully) technical difficulties, we did not manage to complete this calculation so far. We hope to report it later in a separate publication.
In section~\ref{sec:loc} we present an analogous calculation
for the critical string both to illustrate the idea and because we find this calculation instructive on its own.  This calculation leads us to an integral which exhibits the localization property, {\it i.e.} it is 1-loop exact. As we explain in section~\ref{sec:loc} there are several reasons to expect that there is a localization story underlying also the general case of  gravitational dressing/$T\bar{T}$ deformation/JT gravity.

It will be very satisfactory to have a derivation of this sort. However, already at this point it would be very puzzling if gravitational dressing and $T\bar{T}$ turned out to be different deformations.
Not surprisingly, identification of the two sheds light on both of them.  On the one hand, the $T\bar{T}$ description provides the raison d'$\hat{\mbox{\rm e}}$tre for the gravitational dressing of an arbitrary quantum field theory---existence of the 
 $T\bar{T}$ operator and its factorization properties proven in \cite{Zamolodchikov:2004ce}  rely neither on the conformal invariance nor on integrability. On the other hand, the $S$-matrix description of gravitational dressing serves as a rigorous  definition of what $T\bar{T}$ deformation is. Indeed, the operator approach of \cite{Smirnov:2016lqw,Cavaglia:2016oda}  is not rigorously justified in a situation when the RG flow is defined by an irrelevant deformation and the resulting theory admits neither a UV fixed point nor well-defined local operators. In fact, this description leaves one wonder whether the resulting theory is UV complete. The $S$-matrix definition eliminates these doubts by providing well-defined and healthy amplitudes at all collision energies.

The rest of the paper is organized as follows. In section~\ref{sec:solution} we solve the JT gravity directly in the flat space. We start with identifying  the proper 2d Poincar\'e symmetry of the resulting $S$-matrix. This allows to identify also the corresponding dynamical physical coordinates--- ``clocks and rods"---of the JT gravity.
 Then we show how the exact dressing formula follows from the  transformation
 explicitly implementing these clocks and rods as the physical coordinates. Finally we demonstrate how the $T\bar{T}$ operator arises at the leading order in  perturbation theory.
 
 In section~\ref{sec:ads} we provide two alternative derivations of the dressed  scattering, both motivated by 
the $NAdS_2$ holography.  In subsection~\ref{sec:flat} we provide another flat space derivation, which is strongly guided by the $NAdS_2$ calculations. In subsection~\ref{sec:schwarz} we derive the dressing by taking the flat limit of  Schwarzian correlators.
 Both derivations are somewhat subtle and exhibit certain ambiguities, which would be a bit challenging to fix without knowing the expected answer. Perhaps, they should be considered as an exercise on how to take the flat space limit of boundary correlators, rather than bona fide independent derivations. 
In fact, we did learn from these derivations two somewhat surprising lessons.

First, in order to reproduce the unitary $S$-matrix (\ref{dressing}) as a flat space limit of the boundary $NAdS_2$ correlators, the  Schwarzian action needs to be slightly modified. Namely, one needs to perform an additional integration over the total length of the boundary, which results in  averaging over the Schwarzian coupling constant with a certain weight. Without this averaging unitarity is lost in the flat space limit. It will be interesting to understand implications of this for the $NAdS_2$ holography itself.
 
Second, as a byproduct, in subsection~\ref{subsec:single} we obtained   a surprisingly simple and direct relation (\ref{AdSLSZ}) between the flat space $S$-matrix and boundary $AdS$ correlators  for the case of scattering of massive particles.  Results of section~\ref{sec:ads} may be considered as a consistency check of this relation.  Once the validity of (\ref{AdSLSZ}) is firmly established, it will be interesting to study the resulting implications of the boundary OPE for the flat space $S$-matrix.

Section~\ref{sec:loc} is somewhat disjoint from the rest of the paper. We present there the calculation of the finite volume spectrum in the critical string case, which we hope to extend to the general JT case in the future. We present our conclusions and comment on the implications of the presented results for the worldsheet theory of the QCD string in section~\ref{sec:last}.

\section{Solving the Flat Space JT Gravity}
\label{sec:solution}
\subsection{The Exact $S$-matrix From Dynamical Coordinates}
\label{sec:exact}
The goal of the present section is to describe the solution of the flat space ({\it i.e.}, $L=\infty$) JT gravity minimally coupled to an arbitrary matter sector.  The action of this theory is given by
\be
S_{JT+matter}=\int d^2\sigma\sqrt{-g}\l\phi R-\Lambda\r+S_{matter}[\psi,g_{\alpha\beta}].
\ee
We stress that the matter fields, schematically denoted by $\psi$, do not directly couple to the dilaton. This is a different coupling than, for example, in \cite{Callan:1992rs} where matter is coupled to $\phi^{-1}g_{\alpha\beta}$ instead. 
By solution we mean the exact $S$-matrix describing scattering around the vacuum of the JT gravity given by
\begin{gather}
\label{gvac}
g_{\alpha\beta}=\eta_{\alpha\beta}=\l\begin{array}{cc} 0&-1\\
-1&0
\end{array}\r\\
\label{phivac}
{ \phi=-{\Lambda\over 4}\eta_{\alpha\beta}\sigma^\alpha\sigma^\beta+c={\Lambda\over 2}\sigma^+\sigma^-+c}\;,
\end{gather}
where we introduced the light cone coordinates
\[
\sigma^{\pm}={1\over \sqrt{2}}(\sigma^0\pm\sigma^1)\;.
\] 
The zero mode $c$ does not affect the present discussion, but has to be kept track of in the path integral approach of the next section.  To avoid confusion, note that, of course, it is impossible to calculate the exact $S$-matrix of an arbitrary matter sector. Instead, our goal is to calculate how it gets modified due to turning on the JT gravity.  The JT gravity does not bring in any new local degrees of freedom---the JT dilaton plays the role of a Lagrange multiplier, which kills the only candidate propagating mode of the metric. As we will see now the only role of the JT gravity is to provide a dynamical system of coordinates.

The first indication for this comes from the following consideration.
Why does one expect to find a Poincar\'e invariant $S$-matrix for the scattering around the vacuum (\ref{gvac}), (\ref{phivac}), given that the dilaton field in this vacuum has non-trivial space-time dependence?
To see the answer it is convenient to fix the conformal gauge,
\[
g_{\alpha\beta}=e^{2\Omega}\eta_{\alpha\beta}\;.
\]
Then the JT action reduces to 
\[
S_{JT}=\int d\sigma^+d\sigma^-\l4\phi\d_+\d_-\Omega-\Lambda e^{2\Omega}\r\;.
\]
This action is invariant under arbitrary holomorphic and antiholomorphic shifts of $\phi$,
\[
\phi\to \phi+f(\sigma^+)+g(\sigma^-)\;.
\]
We see now that the vacuum  (\ref{gvac}), (\ref{phivac}) is invariant under the combination of the coordinate translations with the Galilean shifts of the dilaton
\bea
\label{galilean}
\sigma^\pm&\to&\sigma^\pm+a^\pm \nonumber\\
\phi&\to&\phi-{\Lambda\over 2}(a^+\sigma^-+a^-\sigma^+)\;.
\eea
Note that, at least for the purpose of the scattering problem, this prescription is well-defined, because the conformal gauge fixing on a plane does not leave any residual gauge freedom if one also imposes $g_{\alpha\beta}\to \eta_{\alpha\beta}$ at infinity. The only conformal transformations preserving this property are those from the Poincar\'e subgroup. The latter is a  physical global
 (rather than gauge) symmetry of the scattering\footnote{It will be interesting to study what is the analogue of the BMS symmetry  in this case.}.

We do not expect to find an analogue of (\ref{galilean}), which  would make it possible to restore a symmetry of the JT vacuum in the $AdS_2$ case. Indeed, the important difference between the $L\to \infty$ JT model considered here and the $NAdS_2$ story is that in the latter case the boundary observables are not invariant under the $AdS_2$ isometries. 
However, given that in section~\ref{sec:ads} we find that the Schwarzian prescription  needs to be extended to obtain a unitary flat space $S$-matrix, it is important to carefully check this expectation.

 Returning to the flat space analysis, the situation is analogous to the worldsheet scattering for the critical string.  There one starts with a long string background
\[
X^0=\sigma^0\;,\;\; X^1=\sigma^1\;,
\]
which is invariant under the combined shift of the worldsheet and target space coordinates. 
Then, using $X^0$ and $X^1$ as physical coordinates, one obtains the worldsheet $S$-matrix described by the dressing
formula (\ref{dressing}), applied to a system of 24 massless free bosons representing the transverse string coordinates \cite{Dubovsky:2012wk}. In the Polyakov description this setup is especially similar to the one encountered in the JT gravity: in addition to transverse fields $X^i$, representing propagating physical modes, one finds an additional topological sector consisting of the Polyakov metric and $X^{0,1}$, whose only role is to provide a dynamical set of coordinates.
A detailed derivation of the worldsheet $S$-matrix in the Polyakov formalism\footnote{Following the idea by Juan Maldacena.} is presented in \cite{Dubovsky:2016cog}. Let us see what happens if one treats the JT gravity in a similar way.

In the conformal gauge the JT field equations for the metric and the dilaton take the following form
\begin{gather}
\label{f++}
\d_+^2\phi-2\d_+\Omega\d_\phi=-{1\over 2}T_{++}\\
\d_-^2\phi-2\d_-\Omega\d_-\phi=-{1\over 2}T_{--}
\label{f--}\\
\label{f+-}
{ \d_{+}\d_-\phi={1\over 2}\l\Lambda e^{2\Omega}+T_{+-}\r}\\
\label{Om}
\d_{+}\d_-\Omega=0\;,
\end{gather}
where the energy-momentum tensor is defined here as
\[
T_{\alpha\beta}=-{2\over\sqrt{-g}}{\delta S\over \delta g^{\alpha\beta}}\;.
\]
With the above boundary conditions for the metric  one finds $\Omega=0$ everywhere.
One treats the system (\ref{f++})-(\ref{Om}) as a set of operator equations. The matter dynamics in $\sigma^\pm$ coordinates remains exactly the same as in the absence of gravity.
Just as in the Polyakov case, dressing arises as a consequence of a coordinate change.
Namely, we introduce new dynamical coordinates defined in the conformal gauge as
\[
X^\pm=2{\d_\mp\phi\over \Lambda}\equiv \sigma^\pm+Y^\pm\;,
\]
where at the last step we separated the vacuum contribution.
The motivation for this choice is that, just as the target space coordinates of a string, $X^\pm$ shift by a constant under the physical Poincar\'e translations {(\ref{galilean})} and reduce to $\sigma^\pm$ at the vacuum.

\begin{figure}[t!]
  \begin{center}
        \includegraphics[height=10cm]{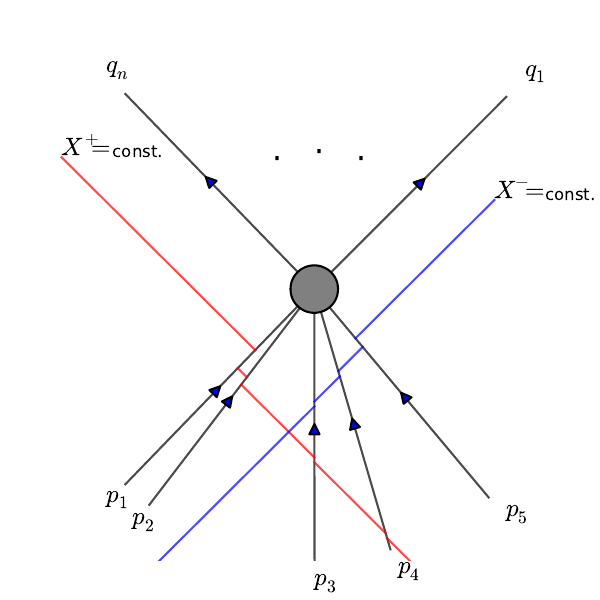} 
           \caption{Lines of constant $X^+$ and $X^-$ as seen in the $\sigma$ coordinates. }
        \label{fig:scatt}
    \end{center}
\end{figure}

Then eqs. (\ref{f++}), (\ref{f--}), (\ref{f+-}) turn into
\begin{gather}
\label{Y-}
\d_+Y^-=-{T_{++}\over \Lambda}\;,\\
\label{Y+}
\d_-Y^+=-{T_{--}\over \Lambda}\;,\\
\label{Yth}
\d_{+}Y^+=\d_{-}Y^-={T_{+-}\over \Lambda}\;.
\end{gather}
Here eqs. (\ref{Yth}) ensure that given $Y^\pm$ one can always find the corresponding dilaton field $\phi$.
As a consequence of the energy-momentum conservation the system (\ref{Y-}), (\ref{Y+}), (\ref{Yth}) always admits a solution. 
This solution is uniquely defined up to a constant shift.
It is natural to pick this constant in a parity symmetric way, so that 
\[
Y^\pm(\sigma^1\to-\infty)=\mp {P^{\pm}\over 2\Lambda}\;,
\]
where
\[
P^\pm=\int_{-\infty}^{\infty} d\sigma^\mp T_{\mp\mp}
\]
are the light cone components of the total 2-momentum carried by the matter.
Then, as a consequence of (\ref{Y-}), (\ref{Y+}),
\[
Y^\pm(\sigma^1\to\infty)=\pm {P^{\pm}\over 2\Lambda}\;.
\]
Consider now a general scattering process taking a set of on-shell momenta $\{p_i\}$ into a set $\{q_j\}$\footnote{Unlike previously, here  we do not treat all momenta as incoming. $\{q_j\}$ is a set of physical momenta without a sign flip.}, see Fig.~\ref{fig:scatt}.
Let us focus on incoming particles, the argument for outgoing ones is identical.
Let us order the incoming momenta by the corresponding rapidities $\beta_i$'s, so that $\beta_1\geq \beta_2\geq\dots\geq\beta_n$.
For in-states this order is equivalent to the order of particles in space.
Then integrating eqs. (\ref{Y-}), (\ref{Y+}) at $\sigma^0\to-\infty$ we find that at early times
\[
Y^{\pm}(\sigma^0\to-\infty,\sigma^1)={1\over 2\Lambda}\l\mp {P^{\pm}}\pm 2P^\pm_L(\sigma)\r\;.
\]
Here $P^\pm_L(\sigma)$ is the total momentum of all particles on the left of $\sigma^1$ at a given $\sigma^0$. This definition is ambiguous when $\sigma$ coincides with a position of one of the particles, $\sigma=\sigma_i$. It is natural to define it there following the central value prescription
\[
P^\pm_L(\sigma_i)={p_i^\pm\over 2}+\sum_{j<i}p_j^\pm\;.
\] 
With this prescription $Y^{\pm}(\sigma_i)$ are independent of $p_i$.
As we will see momentarily, this is very reasonable physically,
because it eliminates contributions into the phase shift which would correspond to particles acting upon themselves.
Explicitly, at the position of  the $i$-th particle these operators are given by
\begin{gather}
\label{Yp}
Y^-(p_i)={1\over 2\Lambda}\l{\cal P}^-_<(p_i)-{\cal P}^-_>(p_i)\r\\
\label{Ym}
Y^+(p_i)={1\over 2\Lambda}\l{\cal P}^+_>(p_i)-{\cal P}^+_<(p_i)\r\;,
\end{gather}
where
${\cal P}^\alpha_<(p_i)$ (${\cal P}^\alpha_>(p_i)$) is an operator which calculates the total momentum of all
particles with smaller (larger) rapidities as compared to $\beta_i$. Note that we replaced the dependence of the operators $Y^\pm$ on $\sigma_i$ coordinates  with the dependence on a particle momentum $p_i$.
This is justified because $Y^\pm$  depend only on the spatial order of particles, which is completely determined by the
rapidities for the in-states.

Before introducing gravity any matter field $\psi$ can be decomposed as
\be
\label{free}
\psi=\int_{-\infty}^\infty {dp \over\sqrt{ 2\pi}}{1\over\sqrt{ 2E}}\l a_{{in}}^\dagger(p) e^{-i p_\alpha\sigma^\alpha}+h.c. \r
\ee
in the asymptotic region $\sigma^0\to -\infty$. 
 
To describe the effect of the JT gravity we need to define the creation operators $A_{in}^\dagger(p)$ using the dynamical coordinates $X^\pm$ rather than $\sigma^\pm$. This amounts to 
\be
\label{dressedA}
A_{in}^\dagger(p)=a_{{in}}^\dagger(p) e^{ip_{\alpha}Y^\alpha(p)}=a_{{in}}^\dagger(p) e^{-{i}\l p^+Y^-(p)+p^-Y^+(p)\r}\;.
\ee
It is straightforward to check that these operators commute
\[
[A_{in}^\dagger(p),A_{in}^\dagger(p')]=0\;,
\]
as it should be for creation operators.
Hence, when creating an in-state, we can put them in an arbitrary order. It is convenient to order them according to the rapidities, so that
\[
|\{p_i\},in\rangle_{dressed}=\prod_{i=1}^{n_{in}}A_{in}^\dagger(p_i)|0\rangle=
e^{-{i\over 2\Lambda}\sum_{i<j}p_i*p_j}|\{p_i\},in\rangle\;.
\]
The argument for  out-states proceeds in exactly the same way, but results in an opposite sign in the final answer for $Y^\pm$
({\it i.e.},  in the analogues of (\ref{Yp}), (\ref{Ym})). Indeed,
out-going particles are {\it antiordered} in space according to their rapidities, which translates into a sign flip in the dressing phase,
\[
|\{q_i\},out\rangle_{dressed}=\prod_{i=1}^{n_{out}}A_{out}^\dagger(q_i)|0\rangle=
e^{{i\over 2\Lambda}\sum_{i<j}q_i*q_j}|\{q_i\},out\rangle\;.
\]
Finally for the dressed $S$-matrix we get
\[
\hat{S}\equiv \;_{dressed}\langle out,\{q_i\}|\{p_i\},in\rangle_{dressed}=\langle out,\{q_i\}|\{p_i\},in\rangle e^{-{i\over 2\Lambda}\sum_{i<j}q_i*q_j}e^{-{i\over 2\Lambda}\sum_{i<j}p_i*p_j}
\]
or, equivalently, it can be written as the operator product
\be
\label{USU}
{\hat S}=USU\;,
\ee
where
\[
U|\{p_i\}\rangle=e^{-{i\over 2\Lambda}\sum_{i<j}p_i*p_j}|\{p_i\}\rangle\;.
\]
As discussed in \cite{Dubovsky:2013ira}, as a consequence of the momentum conservation, (\ref{USU}) is equivalent to 
(\ref{dressing}) provided the relation (\ref{ellL}) holds. Note that in the form (\ref{USU}) the unitarity of the dressed $S$-matrix is manifest, while the initial expression (\ref{dressing}) is explicitly crossing symmetric.
\subsection{Perturbative Scattering and the $T\bar{T}$}
The presented derivation of the $S$-matrix proceeds in a somewhat unconventional way. It is instructive to see how the gravitational scattering arises perturbatively  in a more familiar language. Namely, let us consider the quadratic action for small metric $h_{\alpha\beta}$ and  dilaton perturbations $\varphi$
around the vacuum (\ref{sec:solution}). It takes the following form 
\begin{gather}
S_{JT}^{(2)}=\int \varphi\l\d_+^2h_{--}+\d_-^2h_{++}-2\d_+\d_-h_{+-}\r+{\Lambda\over 4}(h_{++}h_{--}-2h_{+-}^2)\nonumber\\
+{\Lambda\over 4}\l\sigma^+h_{++}(2\d_-h_{+-}-\d_+h_{--})+\sigma^-h_{--}(2\d_+h_{+-}-\d_-h_{++})\r\nonumber\\
+{1\over 2}h_{++}T_{--}+{1\over 2}h_{--}T_{++}+h_{+-}T_{+-}\;.
\label{JTq}
\end{gather}
Metric and dilaton do not contain any propagating degrees of freedom. Hence, we can exclude them using their field equations and this will lead to a local interaction for matter fields.
The  field equations following from the quadratic action (\ref{JTq}) are
\begin{gather}
\d_+^2h_{--}+\d_-^2h_{++}-2\d_+\d_-h_{+-}=0
\label{pheq}
\\
\d_+^2\varphi+{\Lambda\over 4}\l\l 2+\sigma^+\d_+-\sigma^-\d_-\r h_{++}+2\sigma^-\d_+h_{+-}\r=-{1\over 2}T_{++}
\label{h--}
\\
\d_-^2\varphi+{\Lambda\over 4}\l\l 2+\sigma^-\d_--\sigma^+\d_+\r h_{--}+2\sigma^+\d_-h_{+-}\r=-{1\over 2}T_{--}
\label{h++}
\\
\d_+\d_-\varphi +{\Lambda\over 4}\l2h_{+-}-\sigma^+\d_-h_{++}-\sigma^-\d_+h_{--}\r={1\over 2}T_{+-}\;.
\label{h+-}
\end{gather}
A solution to these equations is not unique due to gauge ambiguity. In particular, one can fix the conformal gauge $h_{++}=h_{--}=0$, which also implies $h_{+-}=0$ and brings us to the same situation as before in section~\ref{sec:exact}, where the scattering arises from introducing dynamical coordinates. Instead, here we are looking for an analogue of the static gauge for the Nambu--Goto string, where the worldsheet scattering comes out directly.
The metric solution, which accomplishes this, is 
\be
\label{hT}
h_{\alpha\beta}=-{2\over \Lambda}(T_{\alpha\beta}-\eta_{\alpha\beta} T_\gamma^\gamma)\;.
\ee
To see that this provides a solution, note that as a consequence of the  energy-momentum conservation
\[
\d_+h_{--}-\d_-h_{+-}=\d_-h_{++}-\d_+h_{+-}\;,
\]
which implies also the flatness condition (\ref{pheq}). Furthermore, after plugging (\ref{hT}) in (\ref{h--}), (\ref{h++}), (\ref{h+-}) one arrives at the following set of equations for the dilaton,
\be
\d_\alpha\d_\beta\varphi={1\over 2}(1+\sigma^\gamma\d_\gamma)T_{\alpha\beta}\;.
\ee
These equations are solved by
\begin{gather}
\label{dph+}
\d_+\varphi={1\over 2}(\sigma^+T_{++}-\sigma^-T_{+-})\\
\label{dph-}
\d_-\varphi={1\over 2}(\sigma^+T_{--}-\sigma^+T_{+-})\;.
\end{gather}
It is straightforward to check that (\ref{dph+}), (\ref{dph-}) satisfy the integrability  condition 
\[
\d_+\d_-\varphi=\d_-\d_+\varphi\;.
\]
By plugging (\ref{hT}) back into the action (\ref{JTq}) we find that (at the leading order in $1/\Lambda$) the effect of the JT gravity is equivalent to deforming the matter action by 
\be
 S_{T\bar{T}}=-{1\over 2\Lambda}\int (T_{\alpha\beta}T^{\alpha\beta}-T_\gamma^{\gamma 2})\;,
\ee
which is nothing but the $T\bar{T}$ deformation.

 One can check, using the quadratic part of the stress-energy tensor, that this deformation reproduces the first nontrivial term in the dressing formula \eqref{dressing}. The agreement with the non-perturbative derivation of section \ref{sec:exact} can be understood from the fact that the solution \eqref{hT} is what one would obtain by starting from the conformal gauge and performing a diffeomorphism with parameter $Y^\alpha$ of the previous section. 

\section{Gravitational Dressing as a Flat Limit of $NAdS_2$ holography}
\label{sec:ads}
In this section we establish a link between the $NAdS_2$ holography for the JT gravity and the gravitational dressing of 2D $S$-matrices. We start with a more detailed version of the argument stated in the Introduction that says that the effect of the flat space JT gravity on the $S$-matrix can be expressed as a functional integral similar to (\ref{dressFI}) and present a more careful treatment of this formula. In section \ref{sec:schwarz} we turn to the holographic formulation of the JT gravity in $AdS_2$ given by the Schwarzian dressing (\ref{AdSdressing}). At this point it is natural to expect that  flat space limit of $AdS$ boundary correlators 
will reproduce dressing of the $S$-matrix obtained  directly in flat space. There are several obstacles that complicate  the limit, for example conformal symmetry is broken by the Schwarzian contribution.  
However, with some guidance from the flat space ``holographic" derivation of section~\ref{sec:flat}, we  present a procedure that indeed leads to the anticipated result. Interestingly, this requires a slight modification of the conventional Schwarzian dressing.
\subsection{``Holographic" Derivation of Gravitational Dressing}
\label{sec:flat}
 $S$-matrix is a natural asymptotic observable  in flat space. However, so far there is no well-established procedure to express it purely in terms of a boundary theory. 
 In order to proceed we will consider our theory placed in a finite region of the Minkowski space. The boundary of this region can be roughly thought of as the position of an observer's lab which prepares initial states and measures final ones. To define the $S$-matrix elements we first calculate boundary Greens functions, then take the boundary to infinity and apply the LSZ formula. 

It is convenient to perform the Wick rotation and  to work in the Euclidean flat space{\footnote{We denote all Euclidean variables with a bar and employ the standard  convention for the Wick rotation, $\bar{X}^0=i X^0$, $\bar{X}^1=X^1$.}}. This will prepare us for the $AdS_2$
case, where the Euclidean formulation is almost a necessity, because massive particles do not reach the $AdS_2$ boundary.
 
After setting the metric  to \[ g_{\alpha\beta}=\d_\alpha X^a \d_\beta X^b \eta_{ab}\] and rewriting the area term as a boundary contribution we arrive at the following Euclidean action,
\be
i S_{JT}= - \bar S_{JT}=  -\frac{\Lambda}{2}\oint du \l \dot{\bar{X}}^0 \bar{X}^1-\dot{\bar{X}}^1 \bar{X}^0\r
\;,
\label{CSE}
\ee
where $\bar{X}$'s are the coordinates of the boundary, which parametrize large diffeomorphisms. The boundary curve is fixed by the metric equation of motion to be at $\phi = \phi_b$.

One puzzling aspect of the heuristic argument presented in the Introduction is that the result (\ref{dressFI}) is coming from the vicinity of a trivial saddle point ${\bar X}=0$, which does not have any transparent physical interpretation. On physical grounds one expects the path integral to be dominated by semiclassical configurations corresponding to curves with unit winding. However, the action (\ref{CSE}) does not have any saddle points in sectors with windings: integration over the constant mode of $\phi$ in \eqref{phivac} changes the size of the curve and the area bounded by any non-trivial curve always decreases as the curve shrinks to zero.

This shortcoming can be fixed by adding to the boundary action a counterterm proportional to the length of the boundary,
\be
\label{Sct}
\bar{S}_{ct}=-{\Lambda R_0}\oint du\sqrt{g_{uu}},
\ee
where $g_{uu}$ is the induced metric on the boundary, and $R_0$ is a free parameter. It is convenient to fix reparametrizations along the boundary by requiring the induced metric to be constant,
\be 
g_{uu}={ \delta_{ab} \dot{\bar X}^a\dot {\bar X}^b} = R^2\,,
\label{guu}
\ee
where $R$ is a modulus, which determines the total length of the boundary.  It is directly related to the constant mode of $\phi$.
The background solution for the boundary position in the absence of any matter sources is a circle of radius $R_0$.
This setup is inspired by 
the holographic prescription employed in the $NAdS_2$ case in \cite{Maldacena:2016upp}.

This  ``holographic" formulation  with a dynamical boundary makes it relatively straightforward to understand how asymptotic observables of the original $2d$ QFT are modified in the presence of the JT gravity. Here it is convenient to use the path integral formulation of the theory, rather than the operator language of section \ref{sec:exact}.
The calculation can be done in two steps. We first notice that for any fixed boundary curve the boundary correlators, resulting from the path integral over all matter fields, are identical to those of the original QFT.
Next, we integrate over the boundary fluctuations using the action \eqref{CSE} -- together with the tadpole canceling counterterm (\ref{Sct}) -- to take into account the response of the dilaton to matter fluctuations.
This gives rise to the gravitational dressing. 

It is convenient to introduce radial coordinates 
\be
\bar X^0 = r \cos \theta,\qquad \bar{X}^1 = r \sin\l -\theta\r\;,
\ee
so that the Chern--Simons action (\ref{CSE}) turns into
\be\label{rcs1}
{\bar S}_{JT}=\frac{\Lambda}{2} \oint du r^2 \theta'\;,
\ee
and the condition on the induced boundary metric (\ref{guu}) implies
\be\label{r}
r = \frac{1}{\theta'}\sqrt{R^2-r'^2}.
\ee
Consider small perturbations around the vacuum solution $r=R=R_0$, $\theta = u$,
\be
\label{theta}
\theta = u+\vep(u),
\ee
where 
\[
\vep(0)=\vep(2\pi)\;.
\]
Using \eqref{r} to express perturbations of $r$ through $\vep$ and expanding to second order in perturbations one finds the following expression for the total boundary action
\be\label{svep}
\bar S_{b} = -\pi\Lambda R_0^2+\pi\Lambda (R-R_0)^2 +{\Lambda R^2\over 2} \oint du \l  \vep'^2 -\vep''^2 +O(\vep^3)\r.
\ee
For the purpose of calculating the $S$-matrix we are interested in the $R_0\to \infty$ limit. In this limit $(\Lambda R_0^2)^{-1}$ acts as a weak coupling parameter and the action (\ref{svep}) turns free.

\subsubsection{LSZ Recap}

\begin{figure}[t!]
  \begin{center}
        \includegraphics[height=6cm]{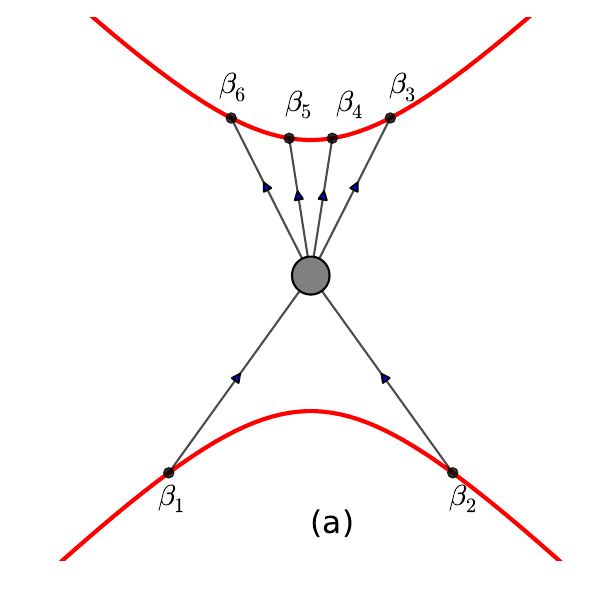} 
        \hspace{-1.cm}
          \includegraphics[height=6cm]{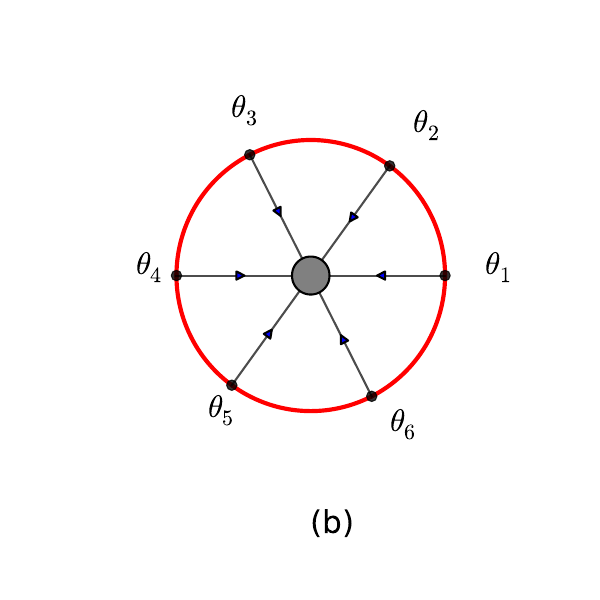} 
        \hspace{-1.cm}
          \includegraphics[height=6cm]{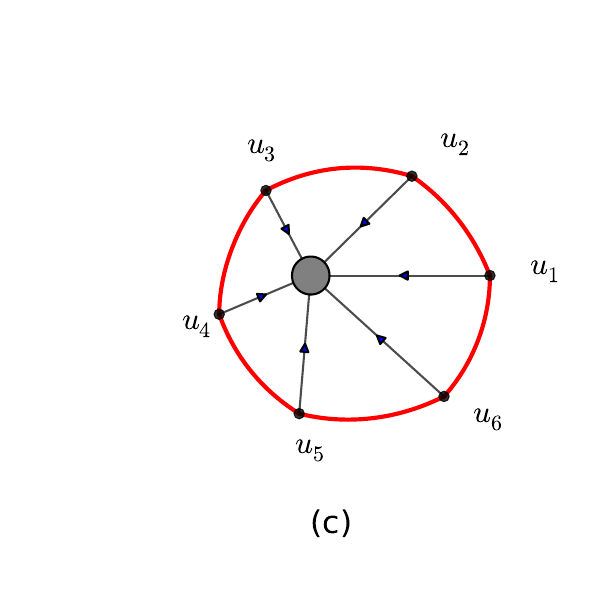} 
           \caption{                (a) Scattering process in the Lorentzian space with LSZ surfaces taken to be hyperbolas, the particles intersect the surfaces at hyperbolic angles determined by their rapidities. (b) The same process but rotated to the Euclidean space. Positions of the particles on the boundary circle are determined by their Euclidean rapidities $\theta_i=\bar \beta_i+\pi$. (c) Coupling to JT gravity results into the deformation of the boundary surface. Positions on the boundary are given by $u_i=\bar \beta_i+\pi$. The grey bulb on all three figures represents the interaction region ${\cal M}_{int}$.}
        \label{fig:Wick}
    \end{center}
\end{figure}

Before proceeding further, let us recall some details of the standard flat space LSZ formalism. It relates the $S$-matrix elements to the time-ordered correlation functions. In its final covariant form, the LSZ prescription requires knowledge of correlators over the entire space-time. However,   to derive the $S$-matrix it is actually enough to know correlation functions of fields inserted on two Cauchy slices in the limit of infinite time separation. Following e.g. \cite{Itzykson}, 
for the $n$-particle scattering amplitude $S_n(\{p_i\})$ we can write\footnote{To keep the equations short, we assume that all wave function normalization constants are set to unity.}
\be
\label{LSZ}
S_n(\{p_i\})=\lim_{T\to\infty}\int \prod_i^n d v_i e^{i\sum_i p_i \cdot X(v_i) }\prod_{i}^n
(n_i\cdot \overset{\leftrightarrow}\d_{X_i})
G_n(\{X(v_i)\}),\ee
where $p_i$ are on-shell momenta, assumed to have negative energies for out-states. $X^\alpha(v)$ belong to the surfaces where in- and out-states are created, the surfaces are parametrized by $v$ and $n_i$ are the corresponding unit normal vectors at points $v_i$, and $T$ is the minimal time separation between in- and out-surfaces. Since both in- and out-fields as well as external wave functions $e^{ip_i \cdot X_i}$ satisfy the Klein-Gordon equation the above expression is independent of the choice of these surfaces. Conventionally, they are taken as two flat constant time hyperplanes. Instead, for our purposes it is convenient to consider hyperbolic spacelike surfaces, which give a connected Euclidean circle of a size $R_0=T$ after the Wick rotation (see Fig.~\ref{fig:Wick}).


Let us  parametrize the momenta via rapidities,
\be
p^0=m \cosh{\beta}, \quad p^1=m\sinh{\beta},
\ee
where for out-states  $\mbox{Im }\beta=\pi$.
To keep the Euclidean path integral real, let us analytically continue rapidities to purely imaginary values $\beta=-i \bar \beta$, which corresponds to $\bar p^0=p^0$ and $\bar p^1= i p^1$. This continuation is distinct from the conventional Wick rotation.
It is however a convenient  analytic continuation of external kinematical data. Since $S$-matrix is an analytic function of rapidities, we are perfectly allowed to calculate it for Euclidean rapidities $\bar\beta$ in the  $(0,2 \pi)$ interval and then analytically continue back into the physical region. We thus arrive at the ``Euclidean" LSZ formula,
\be
\label{LSZE}
 S_n(\{\bar\beta_i\})=\lim_{ R_0\to\infty}\int \prod_i^n( m_i R_0 d\theta_i) e^{- R_0 m_i \cos (\theta_i-\bar\beta_i)} \prod_{i}^n
(n_i\cdot \overset{\leftrightarrow}\d_{X_i}) G_n(\{R_0,\theta_i\}),
\ee
where $ G_n(\{\bar{X}\})$ is the $n$-point function of sources located on the circle of radius $R_0$. 

Note that, as always, in expressions like (\ref{LSZE}) one really has in mind the in- and out- wave functions
to be well localized wave packets rather than true plane waves. These wave packets can be arranged in such a way that all interactions take place in a region ${\cal M}_{int}$ of a finite characteristic size $L_{int}$, such that
\[
p_i^{-1}\ll L_{int}\ll R_0\;.
\]
Then for large $R_0$ the Green's functions can be approximated by expressions of the 
form 
\be
G_n({\bar X_i})\simeq \int_{{\cal M}_{int}} d\bar x \prod_i^n G_2(\bar x-\bar X_i) {\cal K}_n(\{\d_{\bar X_i}\}).
\label{amp}
\ee
The meaning of this schematic formula is that quasilocal kernels ${\cal K}_n(\{\d_{\bar X_i}\})$ determine the momentum structure of the scattering amplitudes, when convoluted with the external wave packets.
For large distances the external propagators take the following asymptotic form
\be
\label{G2}
G_2(\bar x-\bar X_i)= K_0(m_i |\bar x-\bar X_i|)\sim {e^{- m_i |\bar x-\bar X_i|}
\over\sqrt{R_0}} 
\sim {e^{-m_i ({ R_0-\bar X_i\cdot \bar x/R_0})}\over \sqrt{R_0}}\;.
\ee
Keeping only the leading piece in the exponent in (\ref{G2}) one finds that the 
 integral over  $\theta_i$ in \eqref{LSZE} localizes around $\theta_i=\bar\beta_i+\pi$. In the Lorentzian
 signature this localization corresponds to the fact that particles enter the asymptotic region from hyperbolic angles opposite to their rapidities, see Fig.~\ref{fig:Wick}(a). 
 
The subleading pieces in the exponent combine to form 
\[\exp\l\sum_i{-}\l m_i { \cos\bar\beta_i \bar x^0 - m_i\sin\bar \beta_i}\bar x^1\r\r\;,
\]
 which in the Lorentzian signature  turns into $\exp(\sum_i {}i p_i\cdot x)$, ensuring the energy-momentum conservation after integration over $x$. In the Eucledian signature this factor is equal to unity by definition, because the analytic continuation is always performed over a conserved set of momenta.
  Also, at large $R_0$ all the derivatives in the $\prod_in_i\cdot \overset{\leftrightarrow}\d_{X_i}$ factor in  (\ref{LSZE}) can be taken to act on the external wave functions, turning this factor into  $\prod_i - {\bar n}_i\cdot {\bar p}_i\;$. With this, the LSZ formula becomes holographic in the sense that it only depends on correlators restricted to the boundary.

\subsubsection{Gravitational Dressing}
 
Let us come back now to our gravitational problem.
Given the previous discussion the effect of the JT gravity amounts to averaging the Green's functions
over  the positions of the source insertions with the weight determined by the boundary action  (\ref{svep}).
Hence the expression for the dressed $S$-matrix reads
\begin{gather}
 \label{LSZEbig}
\hat {S_n}(\{\bar \beta_i\})=\lim_{R_0\to\infty}\int d\bar x\,dR\, {\cal D} \vep  e^{-\bar S_{b}(\vep,R)}
\\
\nonumber
\prod_i \l { - R_0 d u_i }
e^{-R_0 m_i \cos (u_i-\bar\beta_i)}
{ {\bar n}_i\cdot {\bar p}_i  }
 G_2\left(r(u_i),\theta(u_i),\bar x\right) \r {\cal  K}_n(\{\d_{ \bar X_i}\}).
\end{gather}
Note, that $r(u_i),\theta(u_i)$ now depend on the dynamical variables $\vep,R$ through (\ref{r}), (\ref{theta}).
This rather lengthy expression gets significantly simplified in the large $R_0$ limit. 
Indeed, as we discussed after (\ref{svep}), in this limit $(\Lambda R_0^2)^{-1}$ acts as a small coupling constant, and also the integral over $R$ gets localized in the vicinity of $R_0$. Hence, in this limit the dependence on $\vep$ only survives in the terms where $\vep$ is multiplied either by $R_0$ or $R$.
 Then, using (\ref{G2}),  the  leading part of the external two-point Greens functions reduce to 
 \be
 \label{G2R}
 G_2\sim {e^{-mr}\over \sqrt{ R}}
 \sim {e^{-mR+mR\vep'(u)}\over \sqrt{ R}}\;,
 \ee
 which gives rise to a source for both $\vep$ and $R$.
 As a result we arrive at the following compact expression for the dressed $S$-matrix
 \be
\hat {S}(\{\bar\beta_i\})={\cal N}\int d R e^{-\pi\Lambda (R-R_0)^2}\int {\cal D} \vep e^{-\frac{\Lambda R_0^2}{2}\oint du (\vep'^2-\vep''^2)+\oint du\sum_im_i(R_0\vep'-R+R_0)\delta(u-\bar{\beta}_i-\pi)} S(\{\bar\beta_i\}).
\label{gaussiandr}
\ee
Here, as in the flat case, we took into account that the $\prod du_i$ integral localizes at $u_i=\bar \beta_i+\pi$, and ${\cal N}$ is an overall constant normalization factor.
We see that the integral over $\varepsilon$ is Gaussian and is factored out. We evaluate the path integral over $\vep$ in the Appendix  \ref{app:Y} and after analytic continuation back to real $\beta$  obtain that the dressed
$S$-matrix is given by the expression (\ref{dressing}), up to an additional factor,
\be
\label{Rint}
F=e^{-{\l\sum{m_k}\r^2\over 4\pi\Lambda}}\int d R e^{-\pi\Lambda (R-R_0)^2-\sum m_i( R-R_0)}\;.
\ee
Note that this factor potentially depends on particle masses, so it cannot be set to unity by adjusting the overall constant normalization ${\cal N}$, which should be process independent. 
Fortunately, all the dependence on masses cancels out after evaluating the integral in (\ref{Rint}) and we recover the correct dressing formula  (\ref{dressing}).
%

\subsection{Flat Limit of Schwarzian Correlators}
\label{sec:schwarz}

%
Boundary correlation functions are asymptotic observables in the  $AdS$ gravity.
They are thought to contain all information about a gravitational theory in the bulk. In particular 
it should be possible to extract from them the scattering matrix for states that are localized within {a region} much smaller than the $AdS$ radius.  By taking a family of theories with increasing $AdS$ radii one then expects to extract the full flat space $S$-matrix. This procedure is called the flat space limit of boundary correlators.

In the $AdS_2$ case the (Euclidean) action of the JT gravity takes the following form
\be
\bar S_{JT}= -\int d^2x \sqrt{g} \phi \l R + \frac{2}{L^2}\r  - 2\phi_b\oint_{\cal C} du\sqrt{g_{uu}} K+ 2 c_0\oint_{\cal C} du\sqrt{g_{uu}}.
\label{Adsaction}
\ee
Here we assumed that the space is cut off by some closed boundary ${\cal C}$ with $K$  being its geodesic curvature.
 Variation of the geodesic curvature term w.r.t. the metric together with a boundary variation coming from the bulk JT action
 imply that at the boundary
 \[
 \left.\phi\right|_{\cal C}=\phi_b\;.
 \]
 The last term in (\ref{Adsaction}) is needed to stabilize the position of the boundary, just like in the flat case.
 The cutoff boundary is a usual necessity in the holographic renormalization procedure. However, in a conventional higher dimensional holography no additional boundary dynamics is introduced after the cutoff is removed. The situation is different in the JT $NAdS_2$ case, where gravity affects  correlation functions  exclusively through the boundary dynamics \cite{Almheiri:2014cka}-\cite{Engelsoy:2016xyb}.

Analogously to the flat case the gravitational action reduces to a boundary term and the effect of gravity can be described as dressing of QFT correlators  put in the rigid $AdS_2$ space. The most straightforward way to take the flat space limit of the theory (\ref{Adsaction}) is to move in the boundary ${\cal C}$  deep inside the $AdS_2$ radius. In this way one is guaranteed to eventually reproduce the results of section~\ref{sec:flat}.

On the other hand, if the boundary $AdS_2$ correlators indeed contain all the information about the bulk dynamics, one should 
be able to make the calculation at the $AdS_2$ boundary and 
then perform a transformation of correlators that produces the flat space $S$-matrix. A priori it is not obvious that the two limits commute.

Before proceeding, let us note that we presented the action in the form (\ref{Adsaction}) to agree with the rest of $NAdS_2$ literature. However, one should have in mind that with this choice the flat space limit becomes somewhat obscure. In particular, one may wonder what happened to the vacuum energy $\Lambda$, which used to be the only  parameter in section~\ref{sec:solution}. As pointed out in the Introduction, the answer is related to the possibility to perform a field redefinition,
\be
\label{phishift}
\phi\to\phi+\phi_0\;,
\ee
with a constant $\phi_0$. This results in a simultaneous shift of $\Lambda$ and $\phi_b$,
\[
\Lambda\to \Lambda-{2\phi_0\over L^2}\;,\;\;\phi_b\to\phi_b-\phi_0\;,
\]
so that the actual physical parameter is the invariant combination
\be
\label{lbar}
\bar{\Lambda}=\Lambda-{2\phi_b\over L^2}\;.
\ee
The flat limit $L\to\infty$ is most straightforward if one sets $\phi_b=0$. Nevertheless, in the rest of this section we continue with the choice (\ref{Adsaction}). In the Appendix  \ref{app:F} we  discuss some further details  of the flat space limit, which become more transparent with the choice $\phi_b=0$.

Within the conventional $AdS/CFT$ it is well established how to take the $L\to\infty$ limit resulting in the  scattering of massless particles \cite{Gary,Fitzpatrick:2011jn}. This procedure can be performed directly in the Lorentzian signature and in the global $AdS$ coordinates. However, here we would like to check (\ref{dressing}) also in the massive case. Furthermore, as we discuss in the very end of this section, massless scattering exhibits additional subtleties in the present setup.

Hence, to start with we  consider scattering of massive particles only. This implies that at $L\to\infty$    dimensions $\Delta_a$ of all operators that create scattering states are also taken large, so that  the corresponding masses $m_a=\Delta_a/L$ all  stay finite. 
The flat space limit for massive scattering is subtle even within conventional $AdS/CFT$. 
The main problem is that in the Lorentzian signature massive particles never reach the $AdS$ boundary.
This makes it somewhat challenging to construct scattering states in the boundary theory. 
Hence, similarly to section~\ref{sec:flat}, we work in the Euclidean signature. Our prescription is to  a large extent guided by the flat space  LSZ formalism. 
Morally the resulting  procedure  is similar to the one implemented in  \cite{Paulos}, although  some details are different.
In particular,  we do not require conformal invariance of boundary correlators. 
Once  the massive $S$-matrix is obtained, one can take masses to zero to study the dressing of massless $S$-matrices.
In this sense  mass can be thought of  as an IR regulator. 

\subsubsection{Boundary Action}

In flat space all background solutions for the dilaton are equivalent. This is no longer the case in $AdS_2$. To describe the $AdS_2$  dilaton solutions it is convenient to introduce embedding coordinates $\bar X^A$, satisfying
\be
\label{hyper}
-\bar X_{-1}^2+\bar X_0^2+\bar X_1^2=-L^2\;.
\ee
Here,  as before,  we use bars for Euclidean coordinates.
Then the general dilaton solution reads as
\be
\label{ZX}
\phi= Z\cdot \bar X\;,
\ee
where $Z^A$ are integration constants.  Solutions with time-like, space-like or null vectors $Z$ are qualitatively different.
Only the time-like ones survive in the flat limit (see  Appendix \ref{app:F} for details).
Hence,  in what follows we will focus on the solution with $Z_0$ and $Z_1$ equal to zero.
It is convenient to parametrize the embedding coordinates as
\be
\bar X_{-1}=L\cosh \frac{r}{L},\quad \bar X_0=L \sinh \frac{r}{L} \cos\theta,\quad \bar X_1=L\sinh \frac{r}{L} \sin\l -\theta\r.
\label{embedding}
\ee
In these coordinates the metric reads 
\be
ds^2=dr^2+L^2 \sinh^2\l r/L\r d\theta^2
\label{AdSmetric}
\ee
while the unperturbed boundary will be located at constant $r=R_0$. To get  the boundary action in these coordinates we follow exactly the same path as in section \ref{sec:flat}. We keep the metric in the bulk equal to (\ref{AdSmetric}) and assume that the boundary is dynamical and parametrized by two functions $r(u)$ and $\theta(u)$. To gauge fix the reparametrizations of $u$ we set a condition analogous to (\ref{guu}),
\be
g_{uu}=\frac{L^2}{4}e^\frac{2 R}{L}\;.
\label{gaugeAds}
\ee
As before, using this condition one can express $r(u)$ through $\theta(u)$. Perturbatively for large $R$ one gets
\be
r(u)=R-L \log{\theta'(u)}+Le^{-{2R\over L} }\l\theta'(u)^2-2{\theta''(u)^2\over \theta'(u)^2}\r+{ O}(e^{-{4R\over L}})
\label{r_theta}
\ee
In section~\ref{sec:flat}  it was important to keep the zero mode of the metric $R$ dynamical in order to obtain a unitary $S$-matrix.
 We will see that the same is true for the $L\to \infty$ limit of boundary correlators. Consequently, we allow $R$ to fluctuate around its central value $R_0$. Note that this is different from the conventional $NAdS_2$ prescription, where $g_{uu}$ is kept fixed. The coupling constant $\phi_b$  should scale as $e^{R_0\over L}$ in the $R_0\to\infty$
 limit to keep the action finite. So we set it to 
\be
\label{phib}
\phi_b=-\frac{\Lambda L^2 }{2}\cosh\frac{R_0}{L}.
\ee
In addition to reproducing the correct $R_0\to \infty$ asymptotics at fixed $L$, with this choice  by 
 taking the
 $L\to\infty$ limit at fixed $R_0$  one gets
\[
\left.\phi_b\right|_{L\to\infty}=-\frac{\Lambda L^2 }{2}+{ O}(L^0)\;,
\]
independently of $R_0$. Hence, in view of (\ref{lbar}), we expect to match the earlier flat space results.
The role of the last term in the action (\ref{Adsaction}) is to stabilize the value of $R$. The solution $R=R_0$ is obtained with the choice
\[
c_0=-\frac{\Lambda L}{2}\sinh\frac{R_0}{L}\;.
\]
Then in the large $R_0$ limit the boundary action reads
\be
\bar S_b^{AdS}=\pi\Lambda L^2e^{R-R_0\over L}+{\Lambda L^2}e^{R_0-R\over L}\oint du\l {1\over 2}\theta'(u)^2-{3\over 2}\l{\theta''(u)\over\theta'(u)}\r^2+{\theta'''(u)\over\theta'(u)}\r\;.
\label{SbAds}
\ee
In the second piece we recognize the Schwarzian action \eqref{Sch} with $t=\tan (\theta/2)$. Unlike in previous works, however, the coupling constant in front of the Schwarzian is dynamical and is integrated over. 
As before, one can expand around the classical saddle point by writing
\[
\theta(u)=u+\vep(u)\;.
\]
At finite $L$  the boundary theory (\ref{SbAds}) is highly non-linear. Also, one may worry that integrating over $R$ makes it highly non-local\footnote{Interestingly, this type of non-locality---a promotion of a coupling constant into a fluctuating global dynamical variable--- is analogous to the effect expected from Euclidean wormholes \cite{Coleman:1988cy}. 
}.
However, somewhat surprisingly, one may completely get rid of this non-locality by changing the gauge condition. Indeed, instead of the proper time gauge defined by (\ref{gaugeAds}) one may choose to work in the ``static" gauge, defined by fixing
\[
\theta=u\;.
\] 
In this case  the boundary action is manifestly local and at large $R_0$ takes the following form,
\be
\bar S_b^{(s)}=\Lambda L^2\oint du\l \cosh\l\frac{r-R_0}{L}\r-\frac{1}{2}e^{\frac{R_0-r}{L}}r'^2\r,
\ee 
where we now kept $r(u)$ as a dynamical degree of freedom. 
In this form the prescription of \cite{Jensen:2016pah}-\cite{Engelsoy:2016xyb} appears non-local, implying the following integral constraint
\[
\oint du e^{r\over L}= 2\pi e^{R_0\over L}\;.
\]
We checked that, at least as far as the $L\to \infty$ $S$-matrix goes, both gauge conditions give rise to the same results, so we will continue with the initial choice (\ref{gaugeAds}).

In the  $L\to\infty$ limit nonlinearities of the action (\ref{SbAds}) are suppressed  and only the Gaussian piece survives. 
The resulting quadratic action is 
\be
\bar S_b^{(2)}= \pi \Lambda (R-R_0)^2+{\Lambda L^2\over 2}\oint du \l\vep'^2-\vep''^2\r\;.
\label{Sb2AdS}
\ee
This action does look very similar,  up to some normalizations, to the dressing action (\ref{svep}), which we previously obtained directly in the flat  space calculation.
To check whether we indeed reproduce the same flat space dressing from the $AdS$ calculation, we need to find the form of sources, which need to be added to (\ref{Sb2AdS}) in the presence of scattering particles.

Hence, the next step in our analysis is to define the proper smearing of boundary correlators that will produce the $S$-matrix in the $L\to\infty$ limit. The resulting prescription is analogous to the LSZ formula, which computes the $S$-matrix from the asymptotics of the flat space Green's functions. It is universal and with straightforward modifications can be applied also in higher dimensional theories. 

\subsubsection{Single-particle states}
\label{subsec:single}
Let us consider a massive quantum field theory in a rigid $AdS_2$.
We start with constructing one-particle states that after a proper analytic continuation correspond to plane waves in the center of $AdS_2$.  
In order to make similarity with the LSZ formula manifest we first use the language of bulk fields.  
Consider the following state of the bulk theory
\be\label{Vp}
|V(p)\rangle_{AdS} = \oint d\theta (-P\cdot \bar X(\theta))^{\Delta_V} V(\bar X(\theta))|0\rangle_{AdS}\;,
\ee
where the integral is taken over a constant $r$ circle close to the boundary,
$V(\bar X)$ is a bulk field of a mass $m$ dual to a boundary primary operator $V$ with a dimension 
\[
\Delta_V = 1+\sqrt{1+(mL)^2}\simeq mL\;.
\]
Here $P$ is a vector of the form
\be
P= (0,L\cos\bar\beta, -L\sin\bar\beta).
\ee
and  $\bar\beta$ is a rapidity corresponding to an on-shell Euclidean momentum of the form
\[
\bar p^\alpha = (m \cos\bar\beta, m\sin\bar\beta)\;.
\]
As $\bar X(u)$ approaches the boundary of $AdS$, $e^{r \Delta_V } V(\bar X)$ corresponds to 
a boundary operator {${\cal V}(u)$} dual to $V$, and the bulk state (\ref{Vp}) turns  into the following smeared state of the boundary theory,
\be
|V(p)\rangle_{AdS}=\oint du \l{ -}\cos(u-\bar\beta)\r^{\Delta_V}{\cal V}(u)|0\rangle. 
\label{smearing}
\ee 
To see the reason for considering this particular state it is instructive to inspect the corresponding bulk wavefunction
\be
\Psi(\bar Y) = {}_{AdS}\langle 0| V(\bar Y)|V(p)\rangle_{AdS} ,
\ee
in the flat patch close to the center of the $AdS$, {\it i.e.} for 
\[
\bar{Y} = (L,y \cos\alpha,-y\sin\alpha)
\]
 with $y \ll L$.
It is straightforward to check that upon analytic continuation $\bar \beta \to i\beta$ and $\bar Y^0 \to i Y^0$ it takes the form of a plane wave. To verify this, we use the asymptotic behavior of bulk correlation functions, which also determines the form of the bulk-to-boundary correlators,
\be
\expect{V(\bar{Y}) V(\bar{X})}_{AdS} = \frac{1}{(-2 \bar{X}\cdot \bar{Y})^{\Delta_V}}.
\ee
Using (\ref{embedding}) at $r\to\infty$ we get
\be
\label{smally}
\Psi(\bar y) = \oint d\theta \exp\left[\Delta_V \log-\frac{\cos(\theta-\bar\beta)}
{1 - \frac{y}{L} \cos(\theta-\alpha)}\right].
\ee
For the scattering of massive particles, dimensions $\Delta_V$ all grow linearly with $L$.
Then at $\Delta_V\to \infty$ the integral (\ref{smally}) can be evaluated by a saddle point approximation and gets localized at $\theta=\bar\beta+\pi$. 
An analytic continuation of the saddle point answer gives
\be
\label{plane}
\Psi(y) = \sqrt{2\pi\over \Delta_V}e^{i p\cdot y}\;.
\ee
The localization becomes exact in the limit $\Delta_V \to \infty$. 

We see that the situation is very similar to the flat space LSZ formula. Hence, we  follow the same logic. We assume that the wave functions are convoluted with some external wave packets in $P$, that focuses  plane waves in some region small compared to the $AdS$ radius. We keep these external wave packets implicit as they stay smooth in the flat space limit. 
 On the other hand the smearing function in (\ref{smearing}) is important
 both for localization of boundary integrals in the flat space limit and for creating the plane wave states in the center of $AdS$.

\begin{figure}[t!]
  \begin{center}
        \includegraphics[height=10cm]{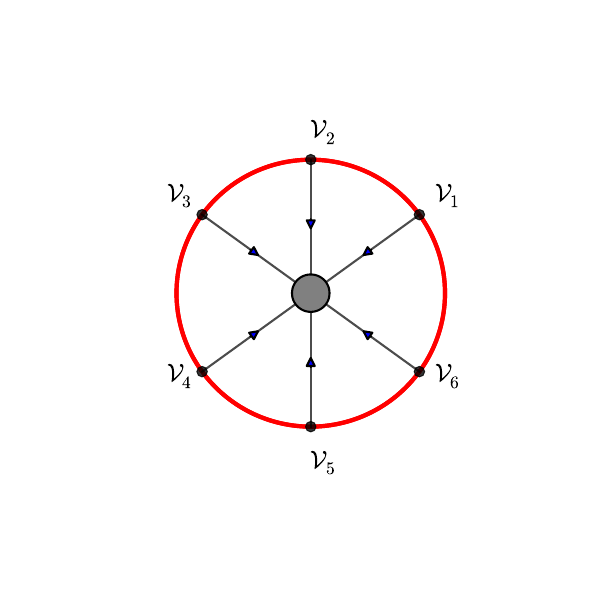} 
           \caption{{The Euclidean $S$-matrix is obtained from the boundary correlators by placing the source corresponding to the $i$-th particle at $\theta_i=\bar{\beta_i}+\pi$.} }
        \label{fig:AdSSmatrix}
    \end{center}
\end{figure}

As a cross-check, we can verify that applying the above construction to a Euclidean correlation function obtained from a Witten diagram with a local bulk vertex reproduces the $S$-matrix which would be generated by the corresponding Feynman diagram. The Witten diagram gives
\be
G(\{\bar X_n\}) = \int_{AdS} d\bar Y \prod_n \frac{1}{(-2 \bar X_n\cdot \bar Y)^{\Delta_{V_n}}} \lambda,
\ee
where $\lambda$ is a coupling constant of the bulk interaction. 
Attaching the state (\ref{Vp}) to each external leg 
 and continuing to the Minkowski momenta and coordinates leads to 
\be
\label{AdSLSZ}
S(\{p_n\}) = i\lambda \int d^{d} y e^{{} i \sum_n p_n\cdot y}= i\lambda \delta^d(\sum_n p_n),
\ee
where we assume analogously to the flat space case that the $y$ integral is taken over some region of $AdS$ space small compared to $L$
and consequently the delta-function is understood to be smoothed over the width of the external wave-packets. More generally, we arrived at the following remarkably simple expression for the Euclidean $S$-matrix\footnote{We thank Shota Komatsu, Joao Penedones and Pedro Vieira for a discussion  on this and for informing us that Shota has independently arrived at this representation \cite{Shota}.} (see Fig.~\ref{fig:AdSSmatrix}),
\be
\label{SO}
S_n\l\{\bar\beta_i\}\r=\lim_{\Delta_i\to\infty}\langle{\Delta_1}^{1/2}{\cal V}_1(\theta_1)\ldots {\Delta_n}^{1/2}{\cal V}_n(\theta_n)\rangle\;, 
\ee
where $\theta_i=\bar\beta_i+\pi$ 
and factors of $\Delta_i^{1/2}$ are introduced to cancel the corresponding prefactor in (\ref{plane}).
It is straightforward to generalize this prescription to higher dimension, where Euclidean rapidities $\bar\beta_i$ should be promoted into  
spherical coordinates of dimensionless Euclidean momenta ${\bar p}^\alpha/m$. It will be interesting to perform further consistency checks on this formula and to derive it from Mellin amplitudes. Here we proceed to calculating the effect of the JT dressing. In fact, this can be considered as one of the consistency checks of (\ref{SO}).

\subsubsection{Dressing}
Calculation of the dressed $S$-matrix  proceeds in the same way as in the Minkowski space. We first place the QFT in question in a rigid Euclidean $AdS_2$ space which produces the set of boundary correlators. We then use the relation (\ref{SO}) between the $S$-matrix and the correlators and
 take the path integral over boundary fluctuations weighted by the action $\bar S_b^{(2)}$, given by equation (\ref{Sb2AdS}), and analytically continue in rapidities. 
The analog of the formula (\ref{LSZEbig}) is then
\be\label{hatSn}
\hat S_n\l\{\bar\beta_i\}\r=\lim_{\substack{L\to\infty\\ \Delta_i/L \to m_i}}\int dR {\cal D} \vep  e^{-\bar S_b^{(2)}}
\prod_i^n\l\l\e^\frac{R_0-R}{L}\theta'(u_i)\r^{\Delta_i}\Delta_i^{1/2}\r 
{ \langle  {\cal V}_1\l\theta(u_1)\r\ldots {\cal V}_n\l\theta(u_n)\r\rangle} \;,
\ee
 where $\theta(u)=u+\vep(u)$ and $u_i\equiv\bar\beta_i+\pi$.
As explained in \cite{Maldacena:2016upp} the rescaling factors
\[
{\cal N}_i=\l\e^\frac{R_0-R}{L}\theta'(u_i)\r^{\Delta_i}
\]
 in front of the correlation function appear from the relation between the bulk fields and operators. Equivalently,  one can derive this factors by looking at the dressed scattering states given by 
\be\label{Vpd}
\widehat{|V(p)\rangle}_{AdS} = \oint du (P\cdot \bar X(u))^{\Delta_V} V(\bar X(\theta(u)))|0\rangle_{AdS}\;,
\ee
where $\bar X(u)$ is given by \eqref{embedding} with $\theta = u$ and $r = R_0$.
${\cal N}_i$'s arise from the mismatch between the boundary asymptotics of $V(\bar{X}(\theta))$ and of $(P\cdot \bar X(u))^{\Delta_V}$ in (\ref{Vpd}) after the use of (\ref{r_theta}). These factors match the conformal transformation rule for boundary primary operators. Hence, 
 in the intrinsic CFT terms dressing operates by averaging over the conformal transformations with the Schwarzian weight (or rather with its extension (\ref{SbAds})).
 In the large $\Delta_i$ limit ${\cal N}_i$ turn into the following source  terms 
\be
\label{AdSsource}
\lim_{\substack{L\to\infty\\ \Delta_i/L \to m_i}}\l\e^\frac{R_0-R}{L}\theta'(u_i)\r^{\Delta_i}=e^{m_i (R_0-R)+L m_i \vep'(u_i)}\;.
\ee
Analogously to the flat case, $\vep(u)$-dependence of the correlator $\langle {\cal V}_1\l\theta(u_1)\r\ldots {\cal V}_n\l\theta(u_n)\r\rangle$ can be neglected at $L\to\infty$, as a consequence of the bulk locality. Namely, the largest sensitivity to boundary fluctuations is in the bulk-to-boundary propagators that are already factored out in \eqref{hatSn}.
Then after performing a rescaling
 \[
 \vep \to {L\over R_0}\vep
 \] 
 the $L\to\infty $  $AdS$ dressing action, given by the sum of (\ref{SbAds})
and (\ref{AdSsource}), exactly matches the flat action in (\ref{svep}). The Gaussian integral over $\vep$ is then the one we computed in the Appendix \ref{app:Y}
so the flat space limit of the Schwarzian dressing indeed reproduces  
 the gravitational dressing (\ref{dressing}). Note, in particular, that if we were to freeze the zero mode of the metric and do not take the integral over $R$ in \eqref{hatSn} we do not get a unitary $S$-matrix.
 
The above derivation relies on the fact that dimensions of operators go to infinity together with the $AdS$ radius. In particular, the source term (\ref{AdSsource}) becomes trivial if one keeps $\Delta_i$'s finite in the flat space limit. 
So, as it stands,  this derivation cannot be applied to massless particles.
 Of course from the results of Section \ref{sec:solution} we know that massless particles in principle can be treated on an equal footing with the massive ones, and that massless scattering can be  obtained as a limit of massive amplitudes. 
 
Massless flat space $S$-matrix is usually associated with the so-called ``bulk-point" singularity
 present in the correlators of holographic CFTs in the Lorentzian signature \cite{Gary}.
 However, there is an additional subtlety in the present setup.
Similarly to the flat space calculation we would like to place a boundary along the lines of constant dilaton. 
Background dilaton solutions which provide constant dilaton slices sufficient to support massless scattering states in the Lorentzian signature {correspond to space-like vectors $Z$ in (\ref{ZX}). Unfortunately, it is straightforward to check that those backgrounds} do not have a smooth flat space limit. 
This is different from the standard situation where the entire $AdS$ boundary is available. 
Nevertheless, massless scattering can be obtained in the limit when energies of particles are much larger than their masses. In this limit the dressing phase (\ref{dressing}) takes a simpler form
\be
\hat S =e^{i \ell^2 E_L E_R} S\;,
\ee
where $E_L$ and $E_R$ are total energies of left- and right-moving particles correspondingly. In particular, for the case of two-to-two massless scattering the dressing factor is equal to $e^{i \ell^2s/4}$ which is a familiar eikonal gravitational phase shift.
\section{Towards Finite Volume Spectrum and Localization}
\label{sec:loc}
So far we established the equivalence between the gravitational dressing and coupling to the JT gravity. We also presented several reasons to expect that this construction is equivalent to the $T\bar{T}$ deformation. Of course, it will be very satisfactory to have a direct derivation of the latter equivalence as well. The JT gravity appears to provide a tractable path integral description for the gravitational dressing. Hence one may hope to derive the generalized Burger's equation (\ref{Burgers}), describing
the $T\bar{T}$-deformed finite volume spectrum, by calculating the toric partition function of the JT gravity coupled to a matter system. Due to several technical difficulties, related mainly to the integration over the overall size of the torus, we did not complete this calculation yet, and hope to present it in a separate publication. Here, instead, we report a prototype calculation, which applies when the matter sector is described by a $c=24$ CFT. Besides illustrating the general idea, this calculation has an independent interest because it gives rise to a localizable ({\it i.e}, one loop exact) integral. We will argue that it is natural to expect the general JT gravity to exhibit the localization property as well.

The $c=24$ conformal matter is special because in this case one can arrive at the dressed $S$-matrix (\ref{dressing}) also by using the conventional Polyakov formalism, similarly to how it happens for the critical bosonic string.
In fact the whole calculation is almost identical to the classic evaluation of the one loop string path integral \cite{Polchinski:1985zf}.
 Namely, instead of introducing a JT sector  one may couple a $c=24$ theory to a metric $g_{\alpha\beta}$ and two massless bosons $X^\mu$, $\mu=0,1$ so that the dressed Euclidean partition function is
\be
Z_{dressed}=\int {{\cal D}g_{\alpha\beta}\over Diff\times Weyl}{\cal D}X^\mu e^{-\int\sqrt{g}{1\over 2\ell^2}(\d_\alpha X^\mu)^2}Z_{CFT}(g_{\alpha\beta})\;,
\ee
where $Z_{CFT}(g_{\alpha\beta})$ is the CFT partition function in the external metric $g_{\alpha\beta}$, and the integral over metrics is moded out by the action of diffeomorphisms and the Weyl group. Note that for a general interacting CFT the $S$-matrix  is an ill-defined object, and it is better to define dressing directly in terms of the finite volume spectrum\footnote{For some CFTs it makes sense to talk about $S$-matrices, understood mostly as a formal object, appearing in the TBA equations which determine the finite volume spectrum, see e.g. \cite{Fendley:1993xa}.}.

For simplicity, to avoid subtleties associated with zero modes, we restrict this discussion to compact CFT's (so that it would need to be modified to cover the first and the simplest example of dressing---the critical bosonic string). The states of the undressed theory put on a circle of size $R$ have energies of the form
\be
\label{CFTen}
E_{n}(R)={2\pi\over R}\l h+\tilde{h}+N+\tilde{N}-{c\over 12}\r\equiv {{\cal E}_n\over R}\;,
\ee
where $h,\tilde{h}$ are conformal weights of the corresponding primaries,  $N$, $\tilde{N}$ are the Virasoro levels, and $n=(h,\tilde{h},N,\tilde{N})$ is a collective label for all of these. The finite volume spectrum of the dressed theory can be extracted from the toric partition function.
To define the latter, recall (see section~\ref{sec:exact}) that the fileds $X^\mu$'s play the role of physical coordinates of the dressed theory in this case. Hence, to calculate the partition function of the dressed theory on a rectangular torus with sides $R_0$ and $R_1$ one needs to integrate over $X^\mu$ field configurations of the form
\be
\label{wind}
X^\mu=R_\mu\sigma^\mu+\delta X^\mu\;,
\ee
where no summation is implied in the first term, and $\delta X^\mu$'s are periodic w.r.t. the worldsheet coordinates $\sigma^\alpha$ with a unit periodicity,
\[
\sigma^\alpha\sim\sigma^\alpha+1\;.
\]
All matter fields also have the same periodicity, and the metric can be chosen in the form
\[
g_{\alpha\beta}d\sigma^\alpha d\sigma^\beta=| d\sigma^0+\tau d\sigma^1|^2\;,
\]
where
\[
\tau\equiv\tau_1+i\tau_2
\]
is the modular parameter. The only difference with the conventional one loop string calculations \cite{Polchinski:1985zf} is that as a consequence of windings for both $X^0$ and $X^1$ in (\ref{wind}) the region of integration over $\tau$ extends to the whole upper half-plane $\tau_2>0$. This is similar to how in \cite{Polchinski:1985zf} the region of integration extends from the fundamental domain of the modular group to the stripe $|\tau_1|<{1\over 2}$, when one moves from the situation with no windings to the one with  a winding of $X^0$ only. Hence, the result in  \cite{Polchinski:1985zf} translates into the following expression for the dressed toric partition function
\be
\label{tauint}
Z_{dressed}(R_0,R_1)={R_0R_1\over\ell^2}\int_0^\infty{d\tau_2\over 2\pi\tau_2^2}\int_{-\infty}^\infty d\tau_1e^{-{Y}}Z_{CFT}(\tau)\;,
\ee
where 
\[
Y={1\over 2\ell^2}\l{R_0^2\over \tau_2}+{R_1^2}\l\tau_2+{\tau_1^2\over\tau_2}\r\r
\]
and $Z_{CFT}(\tau)$ is a toric partition function of the initial CFT. For a compact CFT the latter is given by
\be
Z_{CFT}(\tau)=\sum_ne^{-{\cal E}_n\tau_2-i\tau_1{\cal P}_n}\;,
\ee
where ${\cal E}_n$ are energies of CFT states on a unit circle, (\ref{CFTen}), and
\[
{\cal P}_n=2\pi (N-\tilde{N})
\] 
are the corresponding spatial momenta. At any value of $n$ we end up with an integral of the same form,
\be
\label{In}
I_n={R_0R_1\over\ell^2}\int_0^\infty{d\tau_2\over 2\pi\tau_2^2}\int_{-\infty}^\infty d\tau_1e^{-{Y}}e^{-{\cal E}_n\tau_2-i\tau_1{\cal P}_n}\;.
\ee
This integral is straightforward to evaluate. The integration over $\tau_1$ is Gaussian and leads to 
\be
\label{Bess}
I_n={R_0\over \ell}\int {d\tau_2\over \sqrt{2\pi\tau_2^3}}
e^{-{R_0^2\over2\ell^2\tau_2}-
\tau_2\l {R_1^2\over 2\ell^2}+{\ell^2{\cal P}_n^2\over 2R_1^2}+{\cal E}_n\r\;.
}
\ee
The remaining integration over $\tau_2$ gives
\[
I_n=e^{-{R_0\over \ell}\sqrt{{R_1^2\over \ell^2}+{\ell^2{\cal P}_n^2\over R_1^2}+2{\cal E}_n}}\;.
\]
Hence, the dressed partition function is given by
\[
Z_{dressed}=\sum_n I_n\;,
\]
and the corresponding energy spectrum is
\[
E_{dressed,n}={1\over \ell}\sqrt{{R_1^2\over \ell^2}+{\ell^2{\cal P}_n^2\over R_1^2}+2{\cal E}_n}\;.
\] 
This is exactly the same result as predicted by the generalized Burger's equation (\ref{Burgers}).
Note that this result reproduces correctly also the spectrum of the standard bosonic string, even though  the derivation itself needs to proceed more carefully in that case, to account for the presence of zero modes.  

As we said, we expect that this derivation can be extended to non-critical CFT's and general matter theories by replacing the Polyakov sector with the JT gravity.  In the standard approach to non-critical strings one needs to deal with peculiarities of the Liouville dynamics for the conformal  mode $\Omega$ of the metric, which so far successfully resisted any treatment for $c>1$ matter.  The heuristic argument presented in the Introduction strongly suggests that these problems are mostly ameliorated in the JT
case. Indeed, locally the coupling to the JT sector is equivalent to introducing a $\delta(\d^2\Omega)$ factor in the path integral, which kills most of the Liouville dynamics. 

In the critical case, this factor can be thought of as an alternative way to gauge fix the Weyl symmetry. The presence of $\d^2$ in the argument of the $\delta$-function provides the determinant, which in the standard Polyakov treatment comes from integrating out $X^{0,1}$ modes. In a non-critical/massive case one still needs to deal properly with the integral over the global part of the conformal mode. We leave this task for the future. Here, instead, let us discuss the following interesting aspect of the presented derivation in the critical case.

The brute force derivation above lead us to the integral (\ref{Bess}), which after rescaling of $\tau_2$ and changing an overall constant factor, can be written as 
\be
\label{BesselK}
{1\over 2}\int_0^\infty {d\tau_2\over\tau_2^{\alpha+1}}e^{-{1\over2\hbar}\l\tau_2+\tau_2^{-1}\r}=K_\alpha(\hbar^{-1})
\ee
where $K_\alpha$ is the modified Bessel function, and in our case $\alpha=1/2$, so that we get
\[
K_{1/2}(\hbar^{-1})=\sqrt{\pi \hbar\over 2}e^{-{1\over \hbar}}\;.
\]
At small $\hbar$ the integral (\ref{BesselK}) can be approximated semiclassically, and dominated by the contribution 
 from the vicinity of a saddle point at $\tau_2=1$. The interesting feature of the $\alpha=1/2$ integral is that semiclassics is exact in this case, as if the integral were Gaussian. The answer is exactly given by the semiclassical exponent, multiplied by the corresponding determinant.
 
 Integrals with this property are called localizable. Localization techniques are becoming more and more common in physical applications (see, e.g., \cite{Witten:1992xu,Pestun:2007rz,Pestun:2016zxk,Stanford:2017thb}). Following \cite{Witten:1992xu} many of these applications operate in the context covered by the Duistermaat and Heckman (DH) formula \cite{Duistermaat:1982vw}, {\it i.e.} the corresponding integral is of the form
 \be
 \label{DH}
 \int_{\cal M}\omega^n e^{-\beta H}\;,
 \ee
 where ${\cal M}$ is a symplectic manifold with a symplectic form $\omega$, and the symplectic flow generated by the Hamiltonian $H$ corresponds to the $U(1)$ action on ${\cal M}$.
 
 Let us see that the localization property of the  integral (\ref{BesselK}) with $\alpha=1/2$  can be understood almost in the same way.  Let us first set ${\cal E}_n={\cal P}_n=0$. Then the modular integral we started with, namely
 \be
 \label{I0}
 I=\int_0^\infty{d\tau_2\over \tau_2^2}\int_{-\infty}^\infty d\tau_1e^{-Y}
 \ee
 indeed has a form of the integral over a symplectic manifold---the Poincar\'e half-plane, so that $n=1$ and the symplectic form $\omega$ is associated with the standard Hermitian metric
 \[
 h_P={d\tau\otimes d\bar{\tau}\over\tau_2^2}\;.
 \] 
 The Hamiltonian flow  generated by $Y$ can be written as
 \be
 \label{Mob}
 \dot{\tau}={1\over 2\ell^2}\l R_1^2\tau^2+R_0^2\r\;.
 \ee
 This is not a $U(1)$ flow. However,  compactness of the flow in the DH localization is needed only to establish the existence of a 
 Riemannian metric invariant under the flow (see, e.g., \cite{Stanford:2017thb}). The Poincar\'e half-plane is a K\"ahler manifold,
 and the flow (\ref{Mob})  generates an isometry of the corresponding Riemannian metric. Hence the DH localization applies even though the flow is non-compact. This explains why semiclassics is exact for (\ref{BesselK}). 
 Finally, the integral (\ref{In}) for general ${\cal E}_n,{P}_n$ can be reduced to (\ref{I0}) by rescalings and shifts of the form
 $\tau_2\to a\tau_2$, and $\tau_1\to\tau_1+b\tau_2$.
 
 We believe it is hardly a coincidence that the $T\bar{T}$ deformation lead us to a localizable integral for a critical matter, and expect the localization to persist in the general case as well. There are several indications that the exact answer (\ref{dressing}) is semiclassical in its nature. First, the main physical consequence of the dressing phase shift is that it introduces a time delay, linearly growing with the collision energy. In the Nambu--Goto case this time delay has a very transparent geometrical meaning---the physical length of a string segment is proportional to its energy, hence the time delay. The exact value of the delay can be obtained by solving the classical Nambu--Goto field equations \cite{Dubovsky:2012wk}. Clearly, also in the general JT case one may calculate the exact time delay by downgrading the derivation presented in section~\ref{sec:exact} from the quantum operator language to the classical one. Also the factorization property of the $T\bar{T}$ operator \cite{Zamolodchikov:2004ce}, providing the basis for the $T\bar{T}$ deformation, is highly reminiscent of the large $N$ factorization, which is semiclassical in its nature.
 A further independent support for the semiclassical nature of the JT gravity comes from the recent observation
   \cite{Stanford:2017thb} that even before taking the flat limit the Schwarzian path integral is one loop exact.
   Finally,  yet another surprising empirical evidence for the localization property of the JT gravity is coming from 
   the QCD strings, as will be explained in the next section.
\section{Future Directions and the QCD String}
\label{sec:last}
In this concluding section let us describe a number of open questions raised by the analysis above, and discuss some future directions. To start with, let us explain the promising and somewhat surprising implications
of the results presented here for the dynamics of confining strings (flux tubes) in the planar gluodynamics.

As we already discussed,  gravitational dressing of 24 free massless bosons gives rise to a theory on the worldsheet of a free critical bosonic string. In particular, this theory is invariant under the non-linearly realized target space Poincar\'e symmetry $ISO(1,25)$. Of course, in this case a convenient manifestly covariant path integral description is provided by the Polyakov formalism. However, as proven in \cite{Dubovsky:2015zey}, gravitational dressing of a single massless boson is also invariant under the non-linearly realized target space Poincar\'e symmetry $ISO(1,2)$. Hence, one may wonder whether this system may serve as a starting point for a new type of non-critical strings. Furthermore this setup allows also a natural generalization to $D=4$, which requires introducing an addtional pseudoscalar field 
on the worldsheet. 

Intriguingly, the TBA analysis \cite{Dubovsky:2013gi,Dubovsky:2014fma} of 
the available lattice data in $D=3$ and $D=4$ gluodynamics \cite{Athenodorou:2010cs,Athenodorou:2016ebg,Athenodorou:2017cmw} suggests that confining strings  may be described as deformations of these simple integrable models---leading to the Axionic String Ansatz (ASA) \cite{Dubovsky:2015zey,Dubovsky:2016cog}.
Up until now a theoretical development of the ASA was largely impeded by the absence of a tractable path integral formalism to describe gravitational dressing. Such a formalism is especially called-for in the 
description of the short string (glueball) sector. 
The reformulation of the gravitational dressing via coupling to the JT gravity gives rise to a hope to resolve this problem.
In particular, restricting to $D=3$, we expect that the theory 
\be
\label{3dYM}
S_{3D}=\int\sqrt{-g}\l\phi R+2\ell_s^{-2}-{1\over 2}(\d X)^2\r
\ee
enjoys a non-linearly realized $ISO(1,2)$ symmetry\footnote{In fact, at the classical level one expects to find the non-linearly realized $ISO(1,D-1)$ for $D-2$ massless bosons coupled to the JT gravity. This symmetry survives at the quantum level only for $D=3, 26$.}. It is likely that to identify this symmetry a first order reformulation of the JT gravity as a Poisson sigma model \cite{Verlinde:1991rf,Cangemi:1992bj} should be useful. Earlier discussions of the JT gravity in relation to non-critical strings can be found in \cite{Fradkin:1981dd,Tseytlin:1992ee}.

After the full symmetry is identified it should be possible to evaluate the partition function in the long string sector (in analogy to section~\ref{sec:loc}) and also to extend this analysis to short strings\footnote{This might require adding global degrees of freedom on the worldsheet, similar to NSR sectors in superstings.} (similarly to the critical string case \cite{Polchinski:1985zf}). This may open the path to put on a solid footing the heuristic analysis of the 
$D=3$ glueball spectrum presented in \cite{Dubovsky:2016cog}. A rather surprising outcome of \cite{Dubovsky:2016cog}
is that glueball quantum numbers apparently all can be determined by the semiclassical ansatz, which a priori is expected to work only in the large spin limit. If confirmed, this surprising effectiveness of semiclassics may be  yet another indication that the partition function for (\ref{3dYM}) exhibits the localization property, similarly to section~\ref{sec:loc}.

Naively, one could expect the string tension to be equal to the cosmological constant in the action (\ref{3dYM}). This expectation does not hold---one finds an extra factor of 2 in the action (\ref{3dYM}). Also the sign of the cosmological constant in (\ref{3dYM}) is opposite (negative) to what one may expect for a positive tension string. A possible explanation for this apparent discrepancy is that to reproduce the correct result for the partition function the action (\ref{3dYM}) needs to be supplemented with additional topological (total derivative) terms which do not affect the scattering, but play a role in the partition function calculation. An example of such a term is 
\[
\int\sqrt{-g}\Box\phi\;.
\]
In the analogue of (\ref{tauint}) this term will contribute a factor of $e^{-Y}$, which in the Polyakov case originated from the $(\d X^\mu)^2$ term in the presence of $X^{0,1}$ windings.

Finally, extending the ASA to $D=2$ confining strings suggests that the corresponding  worldsheet theory may be given by the pure JT gravity without matter, perhaps with an additional deformation and/or extension. Confining strings in $D=2$ are topological and understood much better than in higher dimensions \cite{Gross:1992tu,Gross:1993hu,Cordes:1994fc}. However, the local worldsheet description is not completely understood yet. Interestingly, the candidate action described in \cite{Cordes:1994fc} indeed takes the form of  the JT gravity coupled to a topological $\sigma$-model.
 
Coming back to the present setup as a theory of quantum gravity, one may wonder whether obtaining the exact solution of this simple model 
may shed light on numerous profound puzzles present in quantum gravity. Indeed, there is a vast body of literature exploring  black holes (see, e.g., \cite{Callan:1992rs,Susskind:1993if}),
traversable wormholes \cite{Maldacena:2017axo}, baby universes \cite{Rubakov:1996tw} and other beasts in models which appear as rather minimal deformations of the setup studied here.

The exact $S$-matrix (\ref{dressing}) does exhibit a number of unconventional features specific for gravitational theories. These include the novel UV asymptotics, precluding construction of local observables; a time delay
proportional to the collision energy, as expected for Hawking evaporation in two dimensions; coordinate uncertainty principle \cite{Dubovsky:2012wk}; as well as the saturation of the bound on chaos \cite{Maldacena:2015waa}
for a Rindler observer. However, it falls short of delivering the black hole creation and evaporation. For instance, there is no particle production associated with gravitational interactions in this model. 

It remains unclear at the moment whether resolving this deficiency requires a drastic modification of the whole setup or a minor deformation/change of interpretation may be enough. It appears that the required deformation is likely to introduce a direct coupling between the JT dilaton and matter fields. Indeed, the most straightforward way to introduce black holes into the JT gravity is to perform the dilaton dependent  Weyl rescaling of the JT metric
(see, e.g., \cite{Fitkevich:2017izc} for a recent study of the resulting black holes).

The relation to the QCD strings is likely to be useful for clarifying these issues as well. Indeed, lattice simulations of the large $N$ gluodynamics provide an operational and fully non-perturbative
definition of the confining strings.
If the expectation that their worldsheet theory is described by a non-integrable deformation of the JT gravity is correct, this opens a unique and exciting experimental window into the world of quantum gravity\footnote{With the obvious caveat that by an experiment we understand computer simulations here.}. Also from a purely theoretical viewpoint many of the subtleties of quantum gravity (and of two-dimensional dilaton gravity in particular) are related to the proper identification of the physical observables. To a large extent this is the central theme of the current paper---the only effect of the JT sector is to introduce a dynamical system of physical clocks and rods into the system. A relation to the bulk gluodynamics makes it easier to stay on a solid ground and to properly identify the relevant physical observables, such as the glueball spectra and scattering amplitudes.

Let us conclude the paper with a wild speculation. As emphasized in \cite{Dubovsky:2013ira}, a puzzling aspect of the dressing formula (\ref{dressing}) is that it challenges the conventional notion of naturalness. Namely, (\ref{dressing}) provides a simple recipe to introduce a new UV scale $\ell$ into an arbitrary quantum field theory without destabilizing  any unprotected relevant operators. The Lagrangian description for this construction via JT gravity adds another surprising twist to the story. Namely, the JT gravity may be thought of as an extreme example of ``degravitation"---the vacuum energy $\Lambda$ can take an arbitrary value without curving the space-time. Instead, here it manifests itself as the UV deformation of the scattering amplitudes. We  are unable to refrain from reporting a hallucination that extending this story to higher dimension might open a path towards explaining the $M_{weak}\sim\sqrt{M_{Pl}E_{vac}}$ coincidence.
\section*{Acknowledgements}
We thank Ahmed Almheiri, Monica Guica, Guy Gur-Ari, Kristan Jensen, Shota Komatsu, Dima Levkov, Juan Maldacena, Joao Penedones, Riccardo Rattazzi, Slava Rychkov, Sergey Sibiryakov, Eva Silverstein, Pedro Vieira, Xinkang Wang, Edward Witten, and Sasha Zamolodchikov for fruitful  discussions and correspondence. 
We would like thank the organizers and participants of the GGI workshop ``Conformal Field Theories 
and Renormalization Group Flows in Dimensions $d>2$", where this work was initiated, for a very stimulating meeting.
Research at Perimeter Institute is supported by the Government of Canada through Industry Canada and by the Province of Ontario through the Ministry of Economic Development \& Innovation. This work was
supported in part by the NSF CAREER award PHY-1352119. {The work of MM was performed in part at the Aspen Center for Physics, which is supported by National Science Foundation grant PHY-1066293}. 

\appendix
\section{Evaluation of the boundary  path integral}\label{app:Y}
Let us evaluate the  path integral over $\vep$,  which enters in the dressing formula (\ref{gaussiandr}),
\be
I=\int {\cal D }\vep e^{-\frac{\Lambda R_0^2}{2}\oint du (\vep'^2-\vep''^2)+\oint du\sum_im_iR_0\vep'\delta(u-\bar{\beta}_i-\pi)} \;.
\label{gaussianvep}
\ee
This integral is Gaussian, so it can be evaluated exactly by the saddle point approximation. We do not keep track of the overall constant normalization, which will be fixed at the very end by unitarity.
To evaluate the saddle point action we use the equation of motion for $\vep$
\be
\Lambda R_0( \vep''''+ \vep '') =\sum_n m_n \delta'(u_n),\quad u_n=\bar\beta_n+\pi.
\ee
Expanding $\vep$ in terms of solutions of the homogeneous equation in between the sources we find
\be
\vep(u_{n-1}<u<u_{n}) = a_n + b_n u+ c_n e^{i u} + d_n e^{-i u}\;.
\ee
At the sources $\vep''''$ should have a $\delta'$ contribution, implying that $\vep''$ has a discontinuity 
given by
\be 
\vep''(u_n^+)-\vep''(u_n^-)=\frac{m_n}{\Lambda R_0}\;.
\ee
Then we can fix the coefficients by demanding the continuity and $2\pi$ periodicity of $\vep$ and its first and third derivatives.
Let us set $c_1=d_1=0$, then one finds 
\be
\label{abcd}
\begin{split}
a_n = \sum_{k< n} \frac{m_k}{ \Lambda R_0},\quad
b_n = -\sum_{k=1}^{n_{tot}}{m_k\over 2\pi\Lambda R_0},\quad
c_n = \sum_{k< n} -\frac{m_k}{2 \Lambda R_0}e^{-i u_k}, \quad
d_n = \sum_{k< n} -\frac{m_k}{2 \Lambda R_0}e^{i u_k}\;,
\end{split}
\ee
where $n_{tot}$ is the total number of particles.
Note that as a consequence of  energy and momentum conservation the $c_1=d_1=0$ condition is consistent with (\ref{abcd}).
 
Plugging the solution back in the action and using that $u_k-u_n=\bar\beta_k-\bar\beta_n$, we obtain for the integral (\ref{gaussianvep})
\[
I=De^{-{\l\sum{m_k}\r^2\over 4\pi\Lambda}}\;,
\]
where
\be
\begin{split}
D = \exp\l- \frac{\sum_{\bar \beta_k > \bar \beta_ n} m_n m_k\sin(\bar{\beta}_k-\bar \beta_n)}{2 \Lambda}\r.
\end{split}
\ee
After substituting $\bar\beta_n=i\beta_n$ we get an analytic expression for the dressing factor, valid on the entire physical sheet,
\be
\begin{split}
D = \exp\l -i\frac{\sum_{\rm{ Im }\beta_k < \rm{Im } \beta_ n} m_n m_k \sinh(\beta_k-\beta_n)}{2 \Lambda}\r.
\end{split}
\ee
This is just another way of rewriting the dressing factor (\ref{dressing}).
To understand the analytic continuation to the physical domain of real $\beta$'s, let us recall that on the physical sheet the imaginary part of rapidity differences doesn't change the sign  \cite{Dubovsky:2013ira} and consequently the ordering of rapidities stays fixed. In terms of the conventional Mandelstam variables $s_{nk}=m_n^2+m_k^2+2m_n m_k\cosh(\beta_n-\beta_k)$ physical domain is approached from above. Consequently, when both particles are in or both are out, $\beta_k-\beta_n>0$ implies ${\rm Im }\beta_k > {\rm Im }\beta_ n$. Due to the properties of our phase shift, cross terms containing both in- and out-particles cancel out and can be ignored.

\section{Details of the flat space limit}\label{app:F}
Let us  spell out  here elementary geometrical details of taking the $L\to \infty$ limit
of the $AdS_2$ vacuum (\ref{ZX}), (\ref{AdSmetric}) in the JT gravity. As discussed in section~\ref{sec:schwarz}, the $L\to\infty$ limit of the JT action becomes manifestly smooth when one makes use of the field redefinition (\ref{phishift}) to set $\phi_b=0$. Also, it is convenient to perform a shift of the embedding coordinate $\bar{X}_{-1}\to L+\bar{X}_{-1}$ so that the $AdS_2$ hyperboloid also has a smooth $L\to\infty$ limit. After these two redefinitions (\ref{hyper}) is replaced by
\be
\label{Xm1}
\bar{X}_{-1}=\sqrt{L^2+\bar{X}_0^2+\bar{X}_1^2}-L\simeq{\bar{X}_0^2+\bar{X}_1^2\over 2 L}\;.
\ee
At $L=\infty$ this turns into a plane 
\[X_{-1}=0\;.
\]
{The time-like} dilaton solution (\ref{ZX}) with $Z^0=Z^1=0$, $Z^{-1}=Z$ turns into 
\be
\phi=Z(\bar{X}_{-1}+L)+\frac{\Lambda L^2 }{2}\;.
\ee
For this solution to be smooth in the $L\to \infty $ limit one chooses
\[
Z=-{\Lambda L\over 2}+{\phi_0\over L}\;,
\]
where the constant $\phi_0$ is determined from requiring $\phi=0$ at the boundary cutoff surface ${\cal C}$. 
Then at $L\to \infty$ the dilaton solution becomes
\be
\phi=-{\Lambda\over 4}\l\bar{X}_0^2+\bar{X}_1^2\r+\phi_0\;,
\ee
reproducing the flat space result.
\bibliography{bibdil}
\end{document}